\documentclass[12pt]{article}
\usepackage{epsfig,amssymb,amsmath,bbm,psfrag}

\catcode`\@=11
\DeclareMathAlphabet{\mathpzc}{OT1}{pzc}{m}{it}

\newcommand{\insertfig}[2]{\mbox{\epsfxsize=#1cm \epsfbox{#2.eps}}}
\newcommand{\res}{\mathop{\rm Res}}

\font\cmss=cmss12 
\def\1{\hbox{{1}\kern-.25em\hbox{l}}}
\def\bfZ{\relax{\hbox{\cmss Z\kern-.4em Z}}}

\def \be  {\begin{equation}}
\def \ee  {\end{equation}}
\def \ba  {\begin{eqnarray}}
\def \ea  {\end{eqnarray}}
\def \baa {\begin{eqnarray*}}
\def \eaa {\end{eqnarray*}}
\def \bb  {\begin {thebibliography} }
\def \eb  {\end{thebibliography}}
\def \lab #1 {\label{#1}}

\newcommand\re[1]{(\ref{#1})}

\def \matrix #1 {\left(\begin{array}{cc} #1 \end{array}\right)}

\def \tr {\mathop{\rm tr}\nolimits}
\def \res{\mathop{\rm res}\nolimits}
\def \e  {\mathop{\rm e}\nolimits}
\newcommand\lr[1]{{\left({#1}\right)}}

\newcommand \vev [1] {\langle{#1}\rangle}

\newcommand{\as}{\ifmmode\alpha_{\rm s}\else{$\alpha_{\rm s}$}\fi}
\newcommand{\asbar}{\ifmmode\bar{\alpha}_{\rm s}\else{$\bar{\alpha}_{\rm
s}$}\fi}

\newcommand{\ft}[2]{{\textstyle\frac{#1}{#2}}}

\font\cmss=cmss12 
\def\inbar{\,\vrule height1.5ex width.4pt depth0pt}
\def\IC{\relax\hbox{$\inbar\kern-.3em{\rm C}$}}
\def\IZ{\relax{\hbox{\cmss Z\kern-.4em Z}}}
\def\IR{{\hbox{{\rm I}\kern-.2em\hbox{\rm R}}}}

\def\IP{{\hbox{{\rm I}\kern-.2em\hbox{\rm P}}}}
\def\II{\hbox{{1}\kern-.25em\hbox{l}}}

\def\numberbysection{\@addtoreset{equation}{section}
                     \def\theequation{\thesection.\arabic{equation}}}
\numberbysection

\def\n{\ell}

\newbox\lett\newdimen\lheight\newdimen\lwidth
\def\ontop#1#2{\setbox\lett=\hbox{#2}\lheight\ht\lett
\multiply\lheight by 12 \divide\lheight by 10\relax%
\lwidth\wd\lett \multiply\lwidth by 8 \divide\lwidth by 10\relax #2\kern-
\lwidth%
\raise\lheight\hbox{{$\scriptstyle #1$}}\kern.1ex}

\def\XXint#1#2#3{{\setbox0=\hbox{$#1{#2#3}{\int}$}
     \vcenter{\hbox{$#2#3$}}\kern-.5\wd0}}

\begin{document}

\begin{titlepage}

\thispagestyle{empty}

\vspace*{2cm}

\centerline{\large \bf Fine structure of anomalous dimensions in ${\cal N} = 4$ super Yang-Mills
theory}

\vspace{1cm}

\centerline{\sc A.V. Belitsky$^{a}$, G.P. Korchemsky$^{b}$, R.S.
Pasechnik$^{b,a}$}

\vspace{10mm}

\centerline{\it $^a$Department of Physics, Arizona State University}
\centerline{\it Tempe, AZ 85287-1504, USA}

\vspace{3mm}

\centerline{\it $^b$Laboratoire de Physique Th\'eorique\footnote{Unit\'e
                    Mixte de Recherche du CNRS (UMR 8627).},
                    Universit\'e de Paris XI}
\centerline{\it 91405 Orsay C\'edex, France}

\def\thefootnote{\fnsymbol{footnote}}%
\vspace{1cm}

\centerline{\bf Abstract}

\vspace{5mm}

Anomalous dimensions of high-twist Wilson operators in generic
gauge theories occupy a band of width growing logarithmically with
their conformal spin. We perform a systematic study of its fine
structure in the autonomous $SL(2)$ subsector of the dilatation
operator of planar $\mathcal{N} = 4$ superYang-Mills theory which
is believed to be integrable to all orders in 't Hooft coupling.
We resort in our study on the framework of the Baxter equation to
unravel the properties of the ground state trajectory and the
excited trajectories in the spectrum. We use two complimentary
approaches in our analysis based on the asymptotic solution of the
Baxter equation and on the semiclassical expansion to work out the
leading asymptotic expression for the trajectories in the upper
and lower part of the band and to find how they are modified by
the perturbative corrections.

\end{titlepage}

\setcounter{footnote} 0

\newpage

\pagestyle{plain}
\setcounter{page} 1

\section{Introduction}

Wilson operator product expansions (OPE) \cite{Wil69} are the main tools in
analyses of light-cone dominated processes, with deeply inelastic scattering
being the most prominent representative. These expansions underpin the
incoherence of strong-interaction phenomena occurring at different
space-time scales and yield separation of the long- and short-distance
dynamics of the underlying theory of strong interactions.
The former is encoded in hadronic matrix elements of composite Wilson operators
built from elementary QCD fields $X = F_{\mu\nu}, \, \psi, \, \bar\psi$ with
properly contracted color indices to form $SU(3)$ singlets and an arbitrary
number of covariant derivatives ${D}_\mu$. Among these operators there exists
a subclass of the so-called quasipartonic operators which play a distinguished
role and which we shall study in this paper. These operators arise in the OPE
expansion of light-ray operators of the following (schematic) form
\be\label{LR}
X(z_1 n) X(z_2 n) \ldots X(z_L n) = \sum_{N \ge 0} \sum_{{\n}} C^{({\n})}_{N,L} (z_{12},\ldots,
z_{L1}) \mathbb{O}_{N,L}^{({\n})}(0) + \ldots
\ee
where $n_\mu$ is the light-like vector, $n^2_\mu=0$, while the scalar variables $z_i$ define
light-cone coordinates of the fields and the coefficient function $C_{L,N}(z_{12}, \ldots, z_{L1})$
is a homogenous  polynomial in $z_{ij} \equiv z_i-z_j$ of degree $N$. It is tacitly assumed that the
gauge invariance of the nonlocal operator on the left-hand side is restored by inserting Wilson
gauge links between the fields and by appropriately taking traces over the fields' color indices in
corresponding $SU(N_c)$ representations. The quasipartonic operators $\mathbb{O}_{N,L}^{(\n)}(0)$
are built from $L$ quantum fields (not necessary identical) and $N$ light-cone derivatives acting on
them. They belong to the $SL(2)$ sector of the conformal group and the parameters $L$ and $N$ define
their twist and Lorentz spin, respectively. Conformal symmetry allows one to separate quasipartonic
operators into towers with total derivatives like $(n \cdot
\partial)^p \,\mathbb{O}_{N,L}^{(\n)}(0)$. The latter operators are shown by ellipses in the right-hand
side of \re{LR}. It is also known that starting from twist $L=3$, one can construct several
operators with the same quantum numbers $N$ and $L$. The sum over ${\n}$ in the right-hand side of
\re{LR} is meant to enumerate such operators. The physical meaning of this quantum number will
become clear in a moment.

Due to the interaction between partons, e.g., quarks and gluons, the Wilson operators mix under
renormalization. Diagonalizing the corresponding mixing matrix we can construct the operators
$\mathbb{O}_{N,L}^{({\n})}(0)$ in such a way that they have an autonomous scale dependence governed
by a set of anomalous dimensions $\gamma_{N,L}^{({\n})}$. The matrix elements of the Wilson
operators $\mathbb{O}_{N,L}^{({\n})}(0)$ encode information about partonic structure of hadrons
interacting with hard probes, i.e., photons and weak bosons, and their anomalous dimensions control
the scale dependence of various physical quantities. Leading contributions to physical cross
sections of high-energy scattering are associated with twist-two operators ($L=2$), while
power-suppressed effects with higher twists ($L\ge 3$). In particular, the matrix elements of
twist-$L$ operators between the vacuum and a hadron state determine partonic distribution amplitudes
\cite{BroLep79}
\begin{align}\label{WF}
\vev{0|\mathbb{O}^{({\n})}_{N,L}(0) |P}_{\mu} \sim \int_{0}^1 [dx]_L
\varphi_N^{({\n})}(x_1,\ldots,x_L)f(x_1,\ldots,x_L;\mu)\,,
\end{align}
where $x_i$ are the momentum fractions of partons in the hadron with momentum $P_\mu$. Here, the
integration measure includes the momentum conservation constraint $[dx]_L = dx_1\ldots dx_L\,
\delta(1-\sum_{i=1}^L x_i)$, and the weight function $\varphi_N^{({\n})} (x_1,\ldots,x_L)$ is
uniquely defined by the coefficient function $C_{N,L}^{({\n})}$ in Eq.\ \re{LR}. Both sides of this
relation depend on the renormalization scale $\mu$ and the Callan-Symanzik equation for the Wilson
operators can be translated into the evolution equation for the distribution amplitude
$f(x_1,\ldots, x_L;\mu)$. Solving the evolution equation one finds that $f(x_1,\ldots, x_L;\mu)$ can
be decomposed into the sum of {\em different components} enumerated by
${\n}$ each having an autonomous scale dependence%
\footnote{Strictly speaking, this relation only holds in conformal theory. In theory with
nonvanishing beta-function, the $\mu$-dependence is more complicated.}  governed by the anomalous
dimension $\gamma_{N,L}^{({\n})}$
\begin{align}
\label{SOL} f(x_1,\ldots, x_L;\mu) = \sum_{\n}\sum_{N\ge 0} \phi_{N,L}^{({\n})}(x_1,\ldots,x_L)
\lr{\frac{\mu}{\mu_0}}^{-\gamma_{N,L}^{({\n})}}a_{N,L}^{({\n})}{(\mu_0)}\,,
\end{align}
where the functions $\phi_{N,L}^{({\n})}(x_1,\ldots,x_L)$ are orthonormal to the set of the weight
functions in \re{WF}, $\mu_0$ is some reference scale and $a_{N,L}^{({\n})} (\mu_0) =
\vev{0|\mathbb{O}^{({\n})}_{N,L}(0) |P}_{\mu_0}$ is a nonperturbative parameter normalized at
$\mu_0$.

All ingredients of the distribution amplitude in the right-hand side of \re{SOL} are important for
successful phenomenological description of experimental data. Theoretically, however, these are the
anomalous dimensions which represent ``easier'' observables for analytical treatment since they can
be calculated starting from perturbation theory in gauge coupling $g_{\rm\scriptscriptstyle YM}$.
The anomalous dimensions $\gamma_{N,L}^{({\n})}$ are defined as eigenvalues of the mixing matrix
which in its turn represents the dilatation operator of the underlying gauge theory in the
$SL(2;\mathbb{R})$ sector. For given twist $L$, the mixing matrix  can be interpreted as
$L$-particle quantum mechanical Hamiltonian and the problem of finding $\gamma_{N,L}^{({\n})}$ can
be reduced to solving the eigenproblem for this Hamiltonian. The corresponding energy levels are
parameterized by the quantum numbers $N$ and ${\n}$. It is convenient to organize the energy
spectrum into a collection of trajectories labelled by ${\n}$. Each trajectory defines  a smooth
functions of $N$ whose value coincides with $\gamma_{N,L}^{({\n})}$ for $N$ being nonnegative
integer. As in quantum mechanics, the energy levels of the Hamiltonian do not cross and, therefore,
the trajectories with different ${\n}$ and ${\n}'$ do not cross either. This allows us to order them
for each $L$ and arbitrary $N \ge 0$ as
\be\label{order}
\gamma_{N,L}^{(0)} <  \gamma_{N,L}^{(1)} < \ldots \,,\qquad L={\rm fixed} \,,\ N
\ge 0\,,
\ee
where $\gamma_{N,L}^{(0)}$ is the ground state trajectory, $\gamma_{N,L}^{(1)}$ is the first excited
trajectory and so on. We would like to stress that the relation \re{order} holds to any loop order.
In other words, if $\gamma_{N,L}^{(0)} <  \gamma_{N,L}^{(1)}$ at one loop, then the same should be
true for arbitrary coupling. Then, it follows from \re{SOL}, that for $\mu\to\infty$ the sum over
components in the right-hand side of \re{SOL} is dominated by the contribution of the ground
trajectory ${\n}=0$ while the effect of the excited trajectories on the observable is suppressed by
$\mu^{-\Delta_{\n}}$ with $\Delta_{\n} = \gamma_{N,L}^{({\n})}- \gamma_{N,L}^{(0)}>0$. This explains
why the ground state trajectory plays a special role in the analysis of the distribution amplitude
\re{SOL}.

More than this, anomalous dimensions and the properties of the trajectories \re{order} are of
interest in their own right as they were found to reflect a (hidden) symmetry of the underlying
gauge theory. In particular, the one-loop QCD dilatation operator, whose eigenvalues determine
one-loop anomalous dimensions of conformal Wilson operators, was found to possess integrability
symmetry in the so-called aligned-helicity sector. It can be made manifest through the map of the
dilatation operator to the Hamiltonian of an exactly solvable $SL(2)$ Heisenberg magnet
\cite{BraDerMan98,Bel98,BraDerKorMan99,Bel99}. This integrable structure was naturally embedded into
a more general $SU(2,2|4)$ spin chain \cite{Lip97,BeiSta03} spawned by the $\mathcal{N} = 4$
superYang-Mills theory which encodes anomalous dimensions of all single trace operators of this
superconformal theory. The one-loop consideration was extended to all orders in 't Hooft coupling
$g^2 = g_{\rm\scriptscriptstyle YM}^2 N_c/ (4 \pi^2)$ \cite{BeiSta05,Bel08} thus providing a
framework for strong coupling analysis of anomalous dimensions, on the one hand, and a test of the
gauge/string duality between maximally supersymmetric $SU(N_c)$ Yang-Mills theory in the large-$N_c$
limit and the type IIB string theory on AdS$_5 \times$S$^5$ background \cite{Mal97} which relates
anomalous dimensions of Wilson operators to the energies of corresponding string configurations
\cite{GubKlePol02,FroTse02}.

For the twist-two operators, the eigenvalues of the mixing matrix reside on a
single trajectory which is a function of the conformal spin $N$ only. At
large conformal spins the anomalous dimensions displays logarithmic Sudakov
scaling \cite{Kor88,KorMar93},
\be\label{twist2band}
\gamma_{N,L=2} (g) = 2 \Gamma_{\rm cusp} (g) \ln N + \mathcal{O}(N^0) \, ,
\ee
with the overall coefficient determined by the cusp anomalous dimension. For higher twists $L > 2$,
the analytical structure of anomalous dimensions is more complex. In this case the anomalous
dimensions are not specified by the conformal spin alone and form families of nonintersecting
trajectories \re{order} described by the continuous functions $\gamma_{N,L}^{({\n})}$ which depend
on the 't Hooft coupling $g$. At large $N$ the trajectories occupy a band of width which grows
logarithmically with $N$ \cite{BelGorKor03,BelGorKor06},
\be
\label{twistLband} 2 \Gamma_{\rm cusp} (g) \ln N \leq \gamma_{N,L}^{({\n})} (g) \leq L \Gamma_{\rm
cusp} (g) \ln N \, .
\ee
As one approaches the strong coupling regime, one anticipates that these
trajectories do not intercept however their distribution inside the band
gets modified. The minimal trajectory has the same leading large $N$
behavior as for the twist two \cite{BelGorKor03,BelGorKor06}
\be
\gamma_{N,L}^{(0)} (g) = 2 \Gamma_{\rm cusp} (g) \ln N + \mathcal{O}(N^0)
\ee
but subleading $\mathcal{O}(N^0)$ terms are different as compared to
\re{twist2band}.

We would like to stress that the relations \re{twist2band} and \re{twistLband} are valid in any
gauge theory, even in the $SL(2;\mathbb{R})$ sectors which are nonintegrable. The value of the cusp
anomalous dimension depends however on the particle content. In generic (supersymmetric) Yang-Mills
theory, the cusp anomalous dimension is known to two loops \cite{BelGorKor03}, whereas in
$\mathcal{N}=4$ SYM theory it was explicitly calculated in the first four orders of perturbation
theory \cite{BelGorKor03,KotLip02,MocVerVog04,KotLipOniVel04,BerCzaDixKosSmi06}. Later in the paper
we will use its expression to three loops
\be
\label{GammaCusp}
\Gamma_{\rm cusp} (g)
=
\sum_{k = 1} g^{2 k} \Gamma_{{\rm cusp}, k - 1}
=
g^2 - \ft{1}{12} \pi^2 g^4 + \ft{11}{720} \pi^4 g^6
+
\mathcal{O} (g^8) \, .
\ee

The integrability allows one to describe the fine structure of the trajectories.
To one-loop order, the anomalous dimensions in the integrable $SL(2)$ sector
coincide with the energy spectrum of the Heisenberg $SL(2)$ spin chain of
length $L$ and the total spin $N$ and can be found from the Bethe Ansatz.
The Bethe roots take real values only and condense at large $N$ on the union
of intervals on the real axis $[\sigma_1, \sigma_2] \cup [\sigma_3, \sigma_4]
\cup \dots \cup [\sigma_{2 L - 3}, \sigma_{2 L - 2}]$. The values of
$\sigma_i$ can be determined by solving the Bethe Ansatz equations in the
semiclassical approximation. The corresponding solutions are parameterized by
the set of integer numbers
\be\label{ell}
{\n} = \{ {\n}_1,{\n}_2,\ldots,{\n}_{L-1}\} \,,\qquad {\n}_i \ge 0 \,,\quad \sum_{i=1}^{L-1} {\n}_i
=N\,,
\ee
where ${\n}_i$ counts the number of Bethe roots on the interval $[\sigma_{2i- 1},\sigma_{2i}]$ with
$i=1,\ldots,L-1$. The same set of integer numbers determines the anomalous dimension
$\gamma_{N,L}^{({\n})}$ in the semiclassical approach. In order to describe a particular trajectory
in the spectrum of the anomalous dimensions $\gamma_{N,L}^{({\n})}$, it suffices to fix $L-2$
independent integers ${\n}_1,\ldots,{\n}_{L-2}$ and allow $N$ to take positive continuous values.
Going from the lowest to the highest trajectory inside the band \re{twistLband} amounts to counting
all integers ${\n}_i$ satisfying \re{ell}. The one-loop expressions for the trajectories have been
found in \cite{Kor95,BraDerMan98,Bel98,BraDerKorMan99,Bel99} and sum rules for the excited
trajectories to high loops were discussed in \cite{BC}.

The properties of the trajectories depend on both the spin $N$ and twist $L$. At large $N$ the
leading asymptotic behavior of the anomalous dimension $\gamma_{N,L}^{({\n})} \sim \ln N$ does not
depend on the twist $L$. The subleading corrections to $\gamma_{N,L}^{({\n})}$ cease to be
twist-independent and, as was shown in Ref.\ \cite{BelGorKor06}, their properties depend on a hidden
parameter $\xi = \ft1L \ln N$. Namely, for $L,\ N \to \infty$ and $\xi\gg 1$ the minimal anomalous
dimension belonging to the ground state trajectory has the following form
\cite{BelGorKor06,FroTirTse06,AldMal07}
\be\label{GR}
\gamma_{N,L}^{(0)} =\left[ 2\Gamma_{\rm cusp}(g) + \varepsilon(g,j) \right] \ln N +
\mathcal{O}(1/L)\,,
\ee
where $j=1/\xi=L/(\ln N)\ll 1$ and the scaling function $\varepsilon (g,j)$ is given by a double
series in $g^2$ and $j$. From the point of view of the Bethe Ansatz, the asymptotic behavior \re{GR}
corresponds to the configuration of Bethe roots when all but two intervals $[\sigma_{2i - 1},
\sigma_{2i}]$ are shrunk into points. It proves convenient to describe the asymptotic behavior
\re{GR} using a complimentary set of $L$ parameters $\delta_{k}$ (with $k=1,\ldots,L$) given by the
roots of the transfer matrix, the so-called `holes'. Then, the relation \re{GR} corresponds to the
configuration \cite{BelGorKor06} when two holes are large, $\delta_{1}\sim -\delta_{L}\sim N$ while
the remaining $L-2$ `small' holes condense on the interval including the origin of length $\sim
L/\ln N$. In this way, the factor 2 in front of the cusp anomalous dimension on the right-hand side
of \re{GR} just counts the number of `large' holes while the scaling function $\varepsilon(g,j)$
describes the dynamics of `small' holes.

Examining the excited trajectories close to the ground one, one finds that there exists a special
subclass of trajectories for which the leading large-$N$ behavior is $\gamma_{N,L}^{(K)}\sim (K+2)
\ln N$ with $K={\rm fixed}$ as $L,N \to\infty$. In this case, the Bethe roots condense on $K+2$
intervals while the remaining $L-K-1$ intervals collapse into points such
that~\cite{BelGorKor06,Dor08}
\be \label{GR-K}
\gamma_{N,L}^{(K)} = \left[ (K+2) \Gamma_{\rm cusp}(g) +
\varepsilon (g,j_1,\ldots,j_{K+1}) \right] \ln N +
\mathcal{O}(1/L)\,.
\ee
Here  $j_i={\alpha_i}/\xi = {\n}_i L/(N\ln N)$ are scaling variables (with $\alpha_i = {\n}_i/N$
being the filling fractions of Bethe roots on the cuts) obeying $\sum_{i=1}^{K+2} j_i= L/\ln N = j$.
Similar to \re{GR}, the asymptotic behavior \re{GR-K} corresponds in the dual, holes description to
the configuration in which $(K+2)$ holes are large. For different $K$ the scaling functions
$\varepsilon (g,j_1,\ldots,j_{K+1})$ are related to each other through the Whitham flow which
describes a continuous deformation of the distribution density of Bethe roots/`small' holes defined
for different values of the scaling parameters $j_k$~\cite{Kor95,KorKri98}.

We would like to stress that for large $N$  the number of trajectories inside
the band \re{twistLband} scales as $N^{L-2}$ and the relation \re{GR-K} only
holds in the lower part of the band \re{twistLband}. The asymptotic behavior
of the anomalous dimension close to the upper boundary in \re{twistLband} has
a different form
\be
\label{GR-K1} \gamma_{N,L}^{\rm (max)} =   L \Gamma_{\rm cusp}(g)\ln N +
\gamma_L \left(\ln N /N,1/N
\right)\,,
\ee
where the function $\gamma_L$ is given by a series in 't Hooft coupling with the perturbative
coefficients determined in turn by a series in $\ln N/N$ and $1/N$. In this paper, we will compute
$\gamma_L$ for $L=3$ to three loops in ${\cal N}=4$ SYM theory and will demonstrate that the
dependence of $\gamma_L$ on $\ln N/N$ has an interesting iterative structure. Namely, the
perturbative coefficients in front of powers of $\ln N/N$ to higher loops can be expressed in terms
of the cusp anomalous dimension~\cite{BasKor06}
\be\label{g-max}
 \gamma_{N,L}^{\rm (max)} =   L \Gamma_{\rm cusp}(g)\ln \lr{N+\ft12 L
\Gamma_{\rm
 cusp}(g)}+\ldots
\ee
where ellipses denote subleading corrections suppressed by powers of $1/N$. This property can be
made explicit by introducing a new scaling function $f_L(N)$ related to the anomalous dimension via
the recursive relation
\be
\gamma_{N,L}^{\rm (max)} = f_L\left(N+\ft12 \gamma_{N,L}\right)=f_L(N)+\ft14
\lr{f_L^2(N)}'+\ft1{24}\lr{f_L^2(N)}''+
\ldots \, .
\ee
Here the prime denotes a derivative with respect to $N$. We will demonstrate by
explicit calculations
that for $L=3$ the large-$N$ expansion of the function $f_L(N)$ has a form
similar to \re{GR-K1}
with the only important difference that subleading corrections to $f_L(N)$ do
not contain $\ln
N/N^\ell$ terms (with $\ell\ge 1$) to three loops at least. This implies that
$\ln N/N^\ell$
corrections to the anomalous dimension $\gamma_{N,L}^{\rm (max)}$ are induced by
the leading term in
the large $N$ asymptotics of the scaling function $f_L(N) \sim L\Gamma_{\rm
cusp} (g)\ln N +\ldots$.
The same iterative structure has been previously detected for the anomalous
dimension of twist two
\cite{MocVerVog04,DokMarSal06,DokMar06,BasKor06} and for the minimal anomalous
dimension of twist three
\cite{BecDokMar07}, in which case the corresponding scaling function ceases to
be $\ln N/ N^\ell$
independent starting from three loops.

Presently, only the lowest trajectory \re{GR} received thorough studies. As it is evident from Eq.\
\re{GR}, the leading logarithmic asymptotics in the large-$N$ limit allows one to restrict
considerations to the study of the coupling constant dependence of the cusp anomalous dimension. The
latter was found as a solution to an integral equation for the density of Bethe roots
\cite{EdeSta06,BeiEdeSta06} or the resolvent associated with the Baxter function \cite{Bel07}. The
perturbative solutions to the cusp equation were confirmed by field theoretical calculations to
four-loop order in 't Hooft coupling \cite{BerCzaDixKosSmi06}, with the latter providing numerical
confirmation for the conjectured all-order integrable structures in maximally supersymmetric gauge
theory. Many attempts
\cite{BenBenKleSca06,AldAruBenEdeKle07,KotLip07,KosSerVol07,BecAngFor07,CasKri07} to find a
systematic strong coupling solution to the aforementioned equation culminated with the development
of a scheme which allows one to develop inverse coupling expansion \cite{BasKorKot07} (see also
\cite{KosSerVol08}). Due to the strong/weak nature of the AdS/CFT correspondence, this provides a
prediction for the spectrum of corresponding string states. Namely, the anomalous dimension \re{GR}
is related to the energy of a classical folded string spinning with large angular momenta $N$ and
$L$ on the AdS$_3\times {\rm S}^1$ part of the target space \cite{GubKlePol02,FroTse02}. The quantum
corrections to the string sigma-model were calculated to two-loop order \cite{TseRoi08} and were
found to coincide with gauge theory prediction for the cusp anomalous dimension at strong coupling
\cite{BasKorKot07}. In addition, the scaling function $\varepsilon(g,j)$ has been conjectured to be
related to the energy density of the ground state of bosonic O(6) sigma-model~\cite{AldMal07} and
the same result was obtained on the gauge theory side \cite{FreRejSta07,BasKor08}, thus providing
nontrivial dynamical test of the conjectured AdS/CFT correspondence. The string configurations
describing the trajectory \re{GR-K} were suggested to correspond to a string with $K+2$ spikes
approaching the boundary of the anti-de Sitter space \cite{FroTse02,BelGorKor03,Kru05,Dor08} and
rotating with large angular momentum $N$.

So far, no attempts were made however to address trajectories corresponding to
excited states interpolating between the low and upper boundaries of the
spectrum on the gauge theory side, on the one hand, and to find string
configurations dual to them, on the other. Thus the goal of our present
analysis will be to perform the first step in this direction. We will study
the fine structure of the spectrum of high-twist operators inside the band
in the first few lowest orders of perturbative expansion in 't Hooft coupling.
Apart from being of interest to the gauge/string duality, the considerations
done here elucidate an interesting iterative structure of higher twist
anomalous dimensions.

Our subsequent presentation is organized as follows. In the next section we outline our formalism
based on the all-order asymptotic Baxter equation in the $SL(2)$ subsector in the maximally
supersymmetric gauge theory. Then we analyze solutions to the Baxter equation corresponding to
states close to the lower and upper boundaries of the band using two complementary methods,
asymptotic and quasiclassical considerations, respectively. We find that both techniques allow to
describe the fine structure of the spectrum for trajectories going deep into the interior of the
band. They are discussed in Sections \ref{LowerPart} and \ref{UpperPart}, respectively. Finally we
formulate iterative relations for the anomalous dimensions of higher twist operators and then, we
conclude.

\section{Asymptotic Baxter equation}

We focus in this paper on the rank one sector of the $\mathcal{N} = 4$
superYang-Mills theory, which is autonomous to all orders in 't Hooft
coupling constant. It is spanned by single trace operators built from
the complex scalar fields $X$ and covariant derivatives
\be
\label{sl2operators} \mathbb{O}_{N,L} (0) = \sum_{k_1 + {\dots} + k_L = N}
c_{k_1 , {\dots} , k_L}
\tr \, \left\{ \mathcal{D}_+^{k_1} X (0) \dots \mathcal{D}_+^{k_L} X (0)
\right\} \, .
\ee
Here the $+$ subscript stands for the light-cone projection of the corresponding Lorentz index,
i.e., $\mathcal{D}_+ = {D}_\mu n^\mu$ with $n_\mu^2 = 0$. These operators are transformed into each
other under the collinear $SL(2)$ conformal transformation of the four-dimensional conformal
$SO(4,2)$ group.

The operators \re{sl2operators} obey the renormalization group equation
\be
\left( \mu \frac{\partial}{\partial \mu} + \gamma_{N,L} (g) \right)
\mathbb{O}_{N,L} (0) = 0 \, ,
\ee
with anomalous dimensions admitting an infinite loop expansion in
coupling constant,
\be
\label{AnomDimExa} \gamma (g) = \sum_{n = 1}^\infty g^{2 n} \gamma_{n - 1} \, .
\ee
Each term of the series possesses a nontrivial dependence on the quantum numbers $L$ and $N$ and, as
a consequence of the exact solvability, receives an additional dependence on a set of $L-2$ integers
${\n}$ introduced in Eq.\ \re{ell}. Below we review integrable structures arising first at one loop
and then extend them to all orders in 't Hooft coupling within the framework of the Baxter equation.

\subsection{Leading order Baxter equation}

To start with, let us consider the one-loop anomalous dimensions.
Integrability allows one to map the dilatation operator $\mathbb{D} =
\partial/\partial \ln \mu$ into the Hamiltonian of the noncompact
spin-$\ft12$ magnet such that the leading order anomalous dimensions
$\gamma_0$ coincide with its eigenvalues. To find the latter we
resort to the method of the Baxter operator which is known to be
naturally adopted for analyses of spin chains with large values
of quantum numbers.

The central role in the construction is played by the Baxter
$\mathbb{Q}$-operator which depends on the complex spectral
parameter $u$, acts on the spin chain sites and obeys a second
order finite-difference equation. The latter reads for the noncompact
spin-$\ft12$ representations living on the chain sites \cite{Bax82}
\be
\label{LoBaxterEq}
(u^+)^L \mathbb{Q} (u + i) + (u^-)^L \mathbb{Q} (u - i)
= \mathbbm{t} (u) \mathbb{Q} (u) \, .
\ee
Here we introduced
a shorthand notation for the imaginary shifts proportional to the
value of the spins in the spectral parameters $u^\pm \equiv u \pm
\ft{i}{2}$. The right-hand side of the Baxter equation involves a
new operator $\mathbbm{t} (u)$ which is identified as the auxiliary
transfer matrix. It is a polynomial of degree $L$ in the spectral
parameter $u$ with coefficients defined by the local charges
$\mathbbm{q}_k$ acting on the spin chain sites and simultaneously
commuting with the Hamiltonian,
\be \label{tq}
\mathbbm{t} (u) = 2 u^L +
\mathbbm{q}_2 u^{L - 2} + \dots + \mathbbm{q}_L \, .
\ee
The definition of the charges $\mathbbm{q}_k$ is ambiguous since one could have chosen to  expand
$\mathbbm{t} (u)$ around some reference $u=u_0$. A distinguished feature of \re{tq} is that it does
not involve $\mathcal{O}(u^{L-1})$ term, or equivalently $\mathbbm{q}_1=0$. As we will see later,
$\mathbbm{q}_1$ receives perturbative corrections starting from two loops.

The Baxter $\mathbb{Q}$-operator is diagonalized by all eigenstates of the magnet. While the
construction of spin-chain eigenstates requires the operator itself, finding of the energy spectrum
needs only the
knowledge of its eigenvalues which we denote by%
\footnote{We dress the eigenvalue with the subscript $0$ to designate the
fact that we are considering only leading order of the perturbation theory.
Above one-loop, the operator formalism which generalizes the one for the
short-range spin chains based on the existence of $\mathbb{R}$-matrices
is not yet available.}
$Q_0 (u)$. Since the Baxter operator commutes with all local charges, the
equation for its eigenvalues takes the form identical to Eq.\
\re{LoBaxterEq}. This equation however does not fix the form of the
function $Q_0 (u)$ and has to supplemented by the condition on its
analytical properties. In the present setup this boils down to the
polynomiality of $Q_0 (u)$ in the spectral parameter $u$. The large-$u$
asymptotics stemming from the Baxter equation fixes its degree $d=N$ to
be related to the eigenvalues $q_{2,0}$ of the integral of motion
$\mathbbm{q}_2$. The latter is related to the quadratic Casimir operator
of the $sl(2)$ algebra acting on the entire spin chain with eigenvalues
related to the total conformal spin of the chain, $N + \ft12 L$,
\be
q_{2,0} = - (N + \ft12 L) (N + \ft12 L - 1) - \ft14 L \, .
\ee
Being a polynomial of degree $N$,  $Q_0 (u)$ can be parameterized by its roots
$u_{k,0}$ as
\be
\label{LoBaxterEigen} Q_0 (u)
=
\prod_{k = 1}^{N} \left( u - u_{k, 0} \right) \, .
\ee
Since Eq.\ \re{LoBaxterEq} is a homogeneous equation, the overall normalization in Eq.\
\re{LoBaxterEigen} is inessential. The solution to it simultaneously gives the zeroes $u_{k, 0}$ of
$Q_0 (u)$ and quantized values for the conserved charges $q_{k, 0}$. These allow one to determine
the energy, i.e., one-loop anomalous dimensions, and the quasimomentum as
\be
\label{LOgammaANDtheta}
\gamma_0 = \frac{i}{2} \left( \ln \frac{
Q_0 (+ \ft{i}{2}) }{ Q_0 (- \ft{i}{2}) } \right)^\prime
\, , \qquad
\theta_0 = - i \ln \frac{Q_0 (+ \ft{i}{2})}{Q_0 (- \ft{i}{2})}
\, ,
\ee
respectively. The cyclicity of the single trace operators \re{sl2operators} imposes a selection rule
$\theta_0 = 0$ on the quasimomentum.

Substituting the eigenvalues \re{LoBaxterEigen} into the Baxter equation
and taking the residues of both sides at $u = u_{k, 0}$, one immediately
finds that the zeroes of $Q_0 (u)$ obey a set of $N$ transcendental equations
\be
\left( \frac{u^+_{k, 0}}{u^-_{k, 0}} \right)^L = \prod_{j = 1, j \neq k}^N \frac{u_{k, 0} - u_{j, 0}
- i}{u_{k, 0} - u_{j, 0} + i}\,,
\ee
which are identified with Bethe equations.

\begin{figure}[t]
\begin{center}
\mbox{
\begin{picture}(0,200)(220,0)
\put(0,0){\insertfig{6.7}{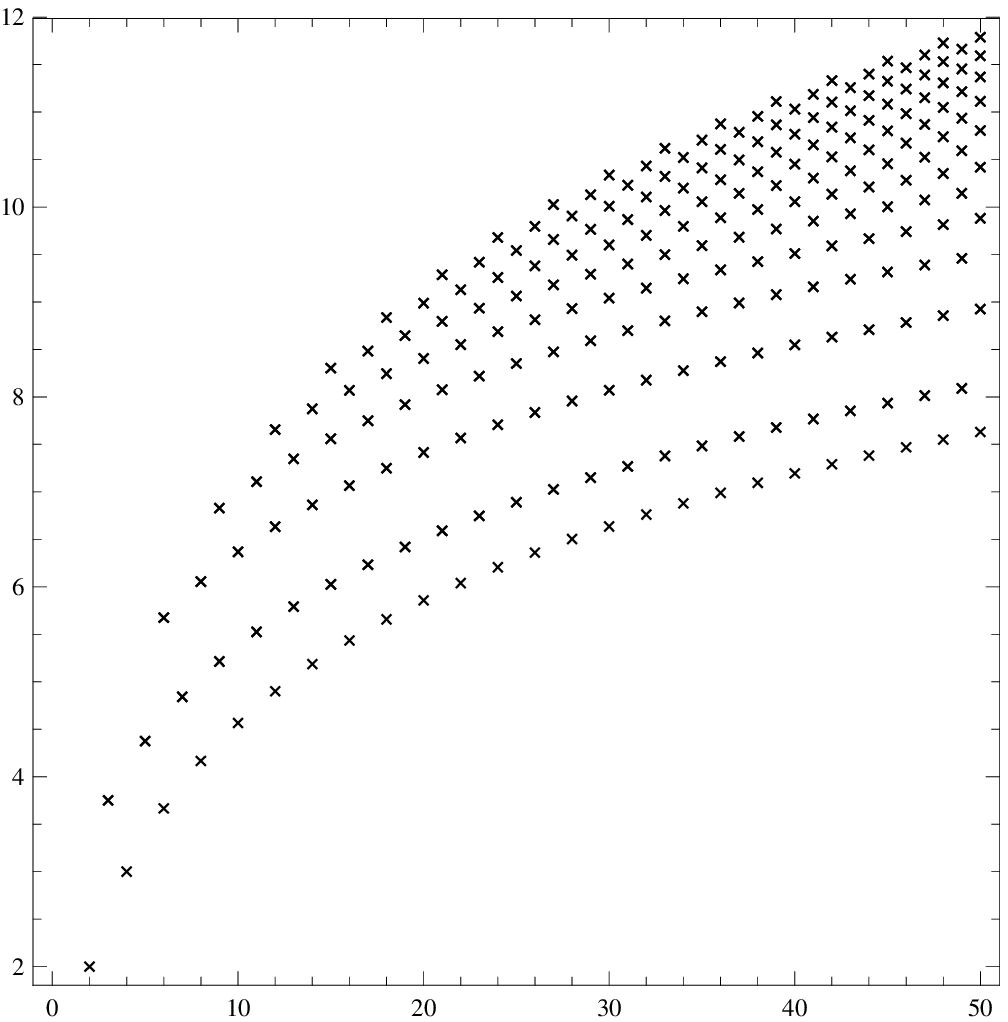}}
\put(-15,98){$\gamma_0$}
\put(170,-10){$N$}
\put(240,0){\insertfig{7}{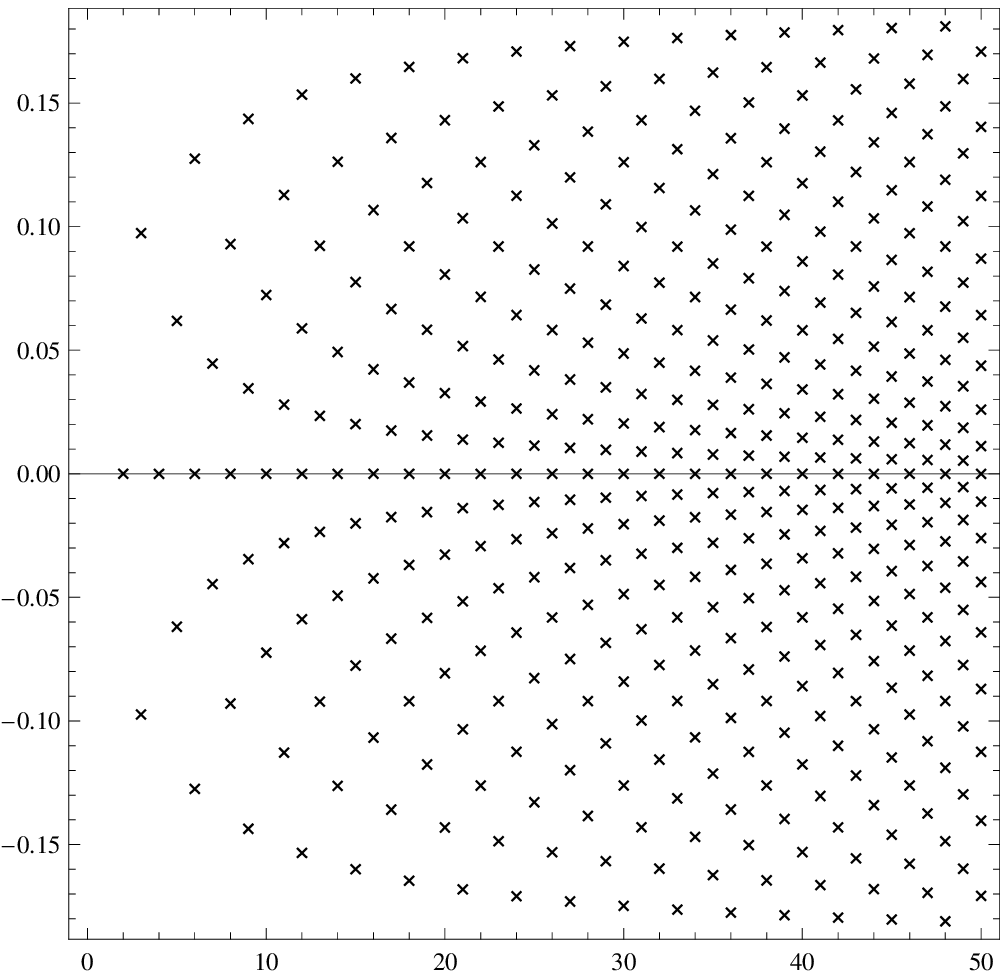}}
\put(215,98){$\widehat{q}_{3,0}$}
\put(415,-10){$N$}
\end{picture}
}
\end{center}
\caption{\label{LOq3ANDgamma} The eigenvalues of the energy $\gamma_0$ and the conserved charge
$\widehat{q}_{3, 0}={q}_{3, 0}/(N+\ft12 L)^3$ for three-site spin chain. For each value of the
conformal spin $N$ there are $m = \ft13 (N-1) + \ft23 {\rm mod}(N-1, 3)$ eigenvalues of the
anomalous dimensions with zero quasimomentum $\theta_ 0 = 0$.}
\end{figure}

The Baxter equation \re{LoBaxterEq} can be solved numerically for fixed values of the parameters $L$
and $N$. We display in Fig.\ \ref{LOq3ANDgamma} the result of such numerical solution for three-site
spin chains for eigenstates with zero quasimomentum and the values of the conformal spin varying in
the range $0 \leq N \leq 50$. The following comments are in order regarding these results. For each
eigenvalue of the energy there are two values of the conserved charge $q_{3, 0}$, so that the former
are double degenerate
\be
\gamma_0 (q_{3, 0}) = \gamma_0 ( - q_{3, 0})
\, ,
\ee
except for the ground state for which $q_{3, 0}$ vanishes. A naked eye inspection unintentionally
traces trajectories in the spectrum in Fig.\ \ref{LOq3ANDgamma}. As it is known from previous
studies \cite{Kor95,BraDerMan98,Bel98,BraDerKorMan99,Bel99}, this is indeed the case as eigenvalues
fall into families of trajectories enumerated by an integer $\ell$. For chains with $L$ sites the
trajectories are parameterized by a set of $L - 2$ numbers $\ell = (\ell_1, \ell_2, {\dots}, \ell_{L
- 2})$.

\subsection{All-order Baxter equation}

Going beyond leading order of perturbation theory one immediately
finds that more than two nearest-neighbor fields simultaneously
interact with each other in the local Wilson operators
\re{sl2operators} such that the dilatation operator becomes
long-ranged. Though, the resulting structures could be potentially
embedded into higher Hamiltonians stemming from the fundamental
transfer matrix of the same short-range $SL(2)$ spin chain, this
simple possibility does not realize. Therefore, one looses operator
formulation of the spectral problem based on the existence of the
$\mathbb{R}$-matrix. Nevertheless, one can consistently deform the
leading order Baxter equation to incorporate higher loop corrections
in 't Hooft coupling. The string-theory considerations imply
that the spectral parameter gets renormalized beyond leading
order of the 't Hooft expansion, i.e., it acquires the coupling
constant dependence \cite{BeiDipSta04},
\be
x [u] = \frac{1}{2} \left( u + \sqrt{u^2 - g^2} \right) \, .
\ee
The construction of the deformed Baxter
equation relies on the function $Q (u)$ which is still polynomial in
$u$ but its roots develop a nontrivial $g$-dependence,
\be Q (u) =
\prod_{k = 1}^N \big( u - u_k (g) \big) \, .
\ee
The resulting equation preserves the form of the second order finite-difference
equation \cite{BelKorMul06,Bel07}
\be
\label{AllOrderBaxterEq}
(x^+)^L {\rm e}^{\Delta_+
(x^+)} Q (u + i) + (x^-)^L {\rm e}^{\Delta_- (x^-)} Q (u - i) = t
(u) Q (u) \, ,
\ee
however with rather involved dressing factors depending on $x^\pm \equiv x
[u^\pm]$,
\be
\Delta_\pm (u) = \sigma_\pm (u) -
\Theta (u) \, .
\ee
Here
\be
\label{Sigmap}
\sigma_\pm (u) =
\int_{-1}^1 \frac{d t}{\pi} \frac{\ln Q (\pm \ft{i}{2} - g
t)}{\sqrt{1 - t^2}} \left( 1 - \frac{\sqrt{u^2 - g^2}}{u + g t}
\right) \, ,
\ee
and  $\Theta (u)$, known as the magnon dressing phase
\cite{AruFroSta04,BeiHerLop06,BeiEdeSta06},
is responsible for the smooth interpolation between the weak- and
strong-coupling expansions
\cite{Bel07b},
\ba
\label{ThetaWeak}
\Theta
(u) \!\!\!&=&\!\!\! g \int_{- 1}^1 \frac{d t}{\sqrt{1 - t^2}} \ln
\frac{Q ( - \ft{i}{2} - g t )}{Q ( + \ft{i}{2} - g t )} \,
{-\!\!\!\!\!\!\int}_{-1}^1 ds \frac{\sqrt{1 - s^2}}{s - t}
\nonumber\\
&\times&
\int_{C_{[i, i \infty]}} \frac{d \kappa}{2 \pi i} \frac{1}{\sinh^2 (\pi \kappa)}
\
\ln
\left( 1 + \frac{g^2}{4 x x [\kappa + g s]} \right)
\left( 1 - \frac{g^2}{4 x x [\kappa - g s]} \right)
\, .
\ea
In complete analogy with the one-loop case, evaluating both sides of the Baxter equation
\re{AllOrderBaxterEq} at $u = u_k (g)$ and taking into account that the renormalized transfer matrix
$t (u)$ is regular at these points one obtains the long-range Bethe Ansatz equations of Ref.\
\cite{BeiSta05}.

For the fixed length $L$ of the operator \re{sl2operators}, the Baxter equation
\re{AllOrderBaxterEq} fails to encode loop corrections above the order $g^{2 (L + 1)}$. When
translated to the range of interaction in the dilatation operator this happens when the interaction
starts to wrap around the chain of fields in the Wilson operator \re{sl2operators}. This is the
so-called wrapping problem which endows the Baxter equation merely with asymptotic character. In the
lack of the operator framework, the functional form of the auxiliary transfer matrix does not appear
to be constrained presently by symmetry considerations. However, the necessity to match the
solutions to the Baxter equation and perturbative results all the way to the wrapping order, i.e.,
including $g^{2 (L + 1)}$, imposes severe constraints on the spectral parameter used as the argument
of $t (u)$. Namely, $t (u)$ admits the form
\be
\label{WrapTransferMat}
t (u) = \Re{\rm e}
\left[
( x^+ )^L
\sum_{k \geq 0} \mathfrak{Q}_k (g) (x^+)^{- k}
\right]
\, ,
\ee
with the sum extending beyond $k = L$. This implies that the transfer matrix
develops non-polynomial effects in the spectral parameter. However, these
are required to precisely cancel the ones emerging from the dressing factors
in the left-hand side of Eq.\ \re{AllOrderBaxterEq}. The expansion
coefficients in Eq.\ \re{WrapTransferMat} are all real, $\Im{\rm m} \,
\mathfrak{Q}_k (g) = 0$ with $\mathfrak{Q}_0 (g) = 2$. Notice also the
appearance of the coefficient $\mathfrak{Q}_1 (g)$ which was absent at
the leading order.

As we just pointed out, the transfer matrix \re{WrapTransferMat} is necessarily
nonpolynomial in the renormalized $x$, and after expansion in 't Hooft coupling
constant, it preserves this virtue also in the bare spectral parameter $u$.
However, if one limits considerations to (and including) order $g^{2 L}$ of
perturbation theory for the anomalous dimensions \re{AnomDimExa}, one finds
that nonpolynomial terms in bare $u$ vanish and the transfer matrix can be
cast in the polynomial form,
\be
\label{PolyTranf}
t (u)
=
u^L \sum_{k = 0}^L q_k (g) u^{- k}
\, ,
\ee
with $q_0 (g) = 2$. The perturbative expansion of charges in both transfer
matrices, Eqs.\ \re{WrapTransferMat} and \re{PolyTranf},
\be
q_k (g)
=
\sum_{n \geq 0}
g^{2 n} q_{k, n}
\, , \qquad
\mathfrak{Q}_k (g)
=
\sum_{n \geq 0}
g^{2 n} \mathfrak{Q}_{k, n}
\, ,
\ee
allows one to establish order-by-order relations between them for $k = 0,
\dots, L$
\ba
q_{k, 0}
\!\!\!&=&\!\!\!
\sum_{m = 0}^{k}
c_{k, m} (L)
\mathfrak{Q}_{m, 0}
\, , \nonumber\\
q_{k, 1}
\!\!\!&=&\!\!\!
\sum_{m = 0}^{k}
c_{k, m} (L)
\left\{
\mathfrak{Q}_{m, 1}
-
\ft14 (L - m + 2)
\theta_{m - 2}
\mathfrak{Q}_{m, 0}
\right\}
\, , \\
q_{k, 2}
\!\!\!&=&\!\!\!
\sum_{m = 0}^{k}
c_{k, m} (L)
\left\{
\mathfrak{Q}_{m, 2}
-
\ft14 (L - m + 2)
\theta_{m - 2}
\mathfrak{Q}_{m, 1}
+
\ft{1}{32}
(L - m + 1)(L - m + 4)
\theta_{m - 4}
\mathfrak{Q}_{m, 0}
\right\}
\, , \nonumber
\ea
with
\be
c_{k, m} (L)
=
\frac{1}{2^{k - m}}
\left(
{L - m \atop L - k}
\right)
\cos \left( \frac{\pi}{2} (k - m) \right)
\, .
\ee
While for $k > L$, one finds
\ba
&&
\mathfrak{Q}_{k > L, 0} = 0
\, , \nonumber\\
&&
\mathfrak{Q}_{L + 1, 1} = \ft14 \mathfrak{Q}_{L - 1, 0}
\, , \qquad
\mathfrak{Q}_{k > L + 1, 0} = 0
\, , \\
&&
\mathfrak{Q}_{L + 1, 2} = \ft14 \mathfrak{Q}_{L - 2, 1}
\, , \qquad
\mathfrak{Q}_{L + 2, 2} = \ft1{16} \mathfrak{Q}_{L - 2, 0}
\, , \qquad
\mathfrak{Q}_{k > L + 2, 2} = 0
\, . \nonumber
\ea

The charges $q_{1} (g)$ and $q_2 (g)$ can be found in a closed form in terms
of the Baxter polynomial $Q (u)$ to all orders of perturbation theory. Using
Eqs.\ \re{Sigmap} and \re{ThetaWeak}, one finds that the first few terms in
the $1/u$-expansion of the dressing factors read
\be
\label{ResduesDressing}
\Delta_\pm (u)
=
\frac{\Delta^{(1)}_\pm (g)}{u}
+
\frac{\Delta^{(2)}_\pm (g)}{u^2}
+
\mathcal{O} (u^{- 3})
\, ,
\ee
where the residues are $(\alpha = 1, 2)$
\be
\Delta^{(\alpha)}_\pm =
\int_{- 1}^1 \frac{d t}{\pi}
\sqrt{1 - t^2}
\left\{
w^{(\alpha)} (g t, g)
\left( \ln Q ( \pm \ft{i}{2} - g t ) \right)^\prime
-
\vartheta^{(\alpha)} (g t, g)
\left(
\ln
\frac{
Q ( + \ft{i}{2} - g t )
}{
Q ( - \ft{i}{2} - g t )
}
\right)^\prime
\right\}
\, ,
\ee
and the explicit form of the functions entering the integrand is
\ba
&&
w^{(1)} (t, g) = - g^2
\, , \qquad
\vartheta^{(1)} (t, g) = 32 i t \frac{\mathcal{Z}_{2,1} (g)}{g^2}
\, , \\
&& w^{(2)} (t, g) = \ft12 g^2 t
\, , \qquad
\vartheta^{(2)} (t, g)
=
8 i \mathcal{Z}_{2,1} (g) \, , \nonumber \ea with \be
\mathcal{Z}_{2,1} (g) = \left( \frac{g}{2} \right)^3
\int_0^\infty dv \frac{J_1 (gv) J_2 (gv)}{v ({\rm e}^{v} - 1)}
\, .
\ee
Then one immediately finds by matching the Taylor series of the left- and
right-hand sides of Eq.\ \re{AllOrderBaxterEq} as $u \to \infty$ that
\ba
\label{q1-def}
q_{1} (g) \!\!\!&=&\!\!\! \Delta^{(1)}_+ (g) + \Delta^{(1)}_- (g)
\, , \\
\label{q2-def}
q_2 (g)
\!\!\!&=&\!\!\!
-
\mathbb{C}_2 (g)
+
\ft14 \left( \Delta^{(1)}_+ (g) + \Delta^{(1)}_- (g) \right)^2
+
\Delta^{(2)}_+ (g) + \Delta^{(2)}_- (g)
-
\ft14 (1 + 2 g^2) L
\, .
\ea
The first term in $q_2 (g)$ is the quadratic Casimir of the $sl(2)$ algebra
\be
\mathbb{C}_2 (g) \equiv (N + \ft12 L + \ft12 \gamma (g))(N + \ft12 L + \ft12
\gamma (g) - 1)
\, ,
\ee
renormalized by the anomalous dimensions of Wilson operators with the conformal
spin
$N + \ft12 L$,
\be
\label{AllOrderAD}
\gamma (g) = - 2 i \left( \Delta^{(1)}_+ (g) - \Delta^{(1)}_- (g) \right)
\, .
\ee
The eigenstates have to be supplemented by the condition of the vanishing
quasimomentum
\be
i \theta
=
\frac{1}{\pi}
\int_{- 1}^1 \frac{dt}{\sqrt{1 - t^2}}
\ln \frac{Q (+ \ft{i}{2} - g t)}{Q (- \ft{i}{2} - g t)}
= 0
\, .
\ee

Finally, before we close this section let us point out yet another (very
suggestive) form of the auxiliary transfer matrix. It allows one to
eliminate the superfluous charge $q_1$ by renormalization of the transfer
matrix $t (u)$ via
\be
t (u) = {\rm e}^{\ft12 [\Delta_+ (x) + \Delta_- (x)]}
\tau (u)
\, ,
\ee
such that
\be
\tau (u) = x^L \left( 2 + \sum_{k \geq 2} \frac{\mathfrak{q}_k (g)}{x^k} \right)
\, ,
\ee
and $\mathfrak{q}_2 (g)$ becomes identical to the eigenvalues of the
renormalized
quadratic $sl (2)$ Casimir operator,
\be
\mathfrak{q}_2 (g) = - \mathbb{C}_2 (g) - \ft14 L
\, .
\ee

\subsection{Analytic structure of the spectrum}
\label{AnalyticStructure}

As we already emphasized in the Introduction, the anomalous dimensions of higher twist operators
occupy a band of width $(L - 2) \Gamma_{\rm cusp} (g) \ln N$ for large conformal spins $N \gg L$.
Numerical analyses demonstrate remarkable regularity of spectra, see, e.g., Fig.\
\ref{LOq3ANDgamma}, of either the energy or the conserved charges. Understanding of these properties
naturally emerges from semiclassical considerations, with the Plank constant played by the parameter
\be
\eta^{- 1} = N + \ft12 L
\, .
\ee
In the small-$\eta$ limit, the Baxter equation reduces to the Shr\"odinger-like
equation for a particle in a singular solvable potential. The size of quantum
fluctuations is suppressed for vanishing $\eta$ and the quantum mechanical
motion
of particles is restricted to their classical trajectories with finite periods.
The classical motion governed by the conserved charges represents a nontrivial
solitonic wave propagating on the chain with $L$ sites. Their quantization stems
from imposing the Bohr-Sommerfeld conditions on periodic classical trajectories.
Thorough studies of leading order Baxter equations have demonstrated that the
eigenvalues of the Hamiltonian (or the anomalous dimension mixing matrix, in the
basis of local Wilson operators) fall onto trajectories parameterized by a set
of integers $\ell$. This property can be understood within the quasiclassical
framework as a consequence of the adiabatic deformation of the solitonic
(finite-gap) solution with continuously varying $N$ which, according to the
adiabatic theorem, will belong to the same trajectory. Since the Hamiltonian
is hermitian, the trajectories do not cross each other as one varies the spin
$N$. The hermiticity of the mixing matrix is preserved beyond one-loop order
of perturbation theory thus one anticipates that these trajectories, while
acquiring nontrivial dependence on the coupling constant, stay non-intersecting
for changing $N$ and $g$. This implies that the numbers $\ell$ remain good
quantum numbers to any loop order and eventually at strong coupling.

It is known \cite{Kor95} that the Bethe roots of the short-range $SL(2)$ magnet with $L$ sites,
which parameterize the zeroes of the Baxter polynomial, admit real values only and populate $(L -
1)$ finite interval on the real axis in the quasiclassical limit $\eta \to 0$,
\be
u \in [\sigma_1, \sigma_2] \cup [\sigma_3, \sigma_4]
\cup \dots \cup [\sigma_{2 L - 3}, \sigma_{2 L - 2}]
\, .
\ee
The leading asymptotic behavior of trajectories in the small $\eta$-limit arises
from configurations when Bethe roots condense on just two intervals on the real
axis. These correspond to the regions of allowed classical motion in the
separated
variables. From the point of view of the spectral curve encoding the analytic
structure of particle quasimomentum, this corresponds to just two open finite
cuts and the rest being collapsed into points, i.e., they do not have any Bethe
roots. The shrunk cuts were dubbed double points in Ref.\ \cite{BelGorKor06}.
The fine structure of the band emerges when these double points open up into
cuts. The single logarithmic asymptotics in the large-spin limit arises when
these two cuts collide at the origin and may or may not pinch the double points
between them. The low boundary of the spectrum corresponds in this picture to
the case when all double points are trapped at the origin by the inner
end-points of the colliding cuts. While the upper boundary of the spectrum is
a consequence of the migration of all these double points outside of the merged
single cut involving the origin. The intermediate states in the spectrum are
described by continuous deformation of the multicut configuration,
when the external cuts gets shrunk with internal cuts opening up at the same
time.

As our discussion suggests, the upper and lower boundaries of the spectrum corresponds to different
analytic structure of the distribution of Bethe roots in the complex plane. These can be translated
into the asymptotic behavior of the conserved charges $q_k (g)$ or, equivalently, the roots
$\delta_k (g)$ of the transfer matrix,
\be
t (u) = 2 \prod_{k = 1}^L \left( u - \delta_k (g) \right)
\, ,
\ee
which similarly to $q_k (g)$ admit perturbative expansion in 't Hooft coupling
constant
\be
\delta_k (g) = \sum_{n = 0} g^{2 n} \delta_{k, n}
\, ,
\ee
and are in one-to-one correspondence to each other,
\begin{eqnarray}
\sum_{k = 1}^L \delta_k (g) = -\ft12 q_1 (g) \, , \quad \sum_{k=1}^L \delta^2_k (g) =\ft14 {q_1^2(g)}-
 q_2 (g) \, ,\  \ldots \quad \prod_{k=1}^L \delta_k (g) = \ft12 (-1)^L q_L (g) \, .
\end{eqnarray}
The upper boundary of the spectrum corresponds to large values of all conserved
charges,
\be
q_k (g) = \eta^{- k} \widehat{q}_k (g)
\, ,
\ee
as $\eta \to 0$ and all rescaled charges $\widehat{q}_k (g)$ being of order one,
$\widehat{q}_k (g) \sim \mathcal{O} (\eta^0)$. The same is applicable to the
roots $\delta_k (g)$. Thus all quantum numbers determining the spin-chain state
are large and quasiclassical approximation is well defined in this regime.

On the other hand, the lower boundary of the spectrum \re{twistLband}
corresponds to the degenerate situation when all higher conserved charges
$q_{k > 2} (g)$ take anomalously small values, as was demonstrated in Ref.\
\cite{BelGorKor06} at leading order of perturbation theory,
\begin{eqnarray}
\widehat{q}_{2k}
\sim
2 \frac{(-1)^k}{(2k)!}\beta^{2k}
+ \mathcal{O} (g^2)
\, ,
\end{eqnarray}
where $ \beta =  {\ft12 L}/\lr{N + \ft12 L} $ vanishes in the single-logarithmic limit $N \gg L$.
This indicates the appearance of two types of roots of the transfer matrix, `large' and `small'
possessing the following asymptotics in the leading order in 't Hooft coupling \cite{BelGorKor06}
\be
\label{scaling} |\delta_{1, L}| > \ft1{\sqrt{2}} N + \mathcal{O} (g^2) \, , \qquad |\delta_{k \neq
1, L} | < \frac{1}{4 \xi} + \mathcal{O} (g^2) \, ,
\ee
which exhibits the emergence of a new ``hidden'' parameter
\be \xi = \frac{1}{L} \ln N
\, . \ee
For the lower part of the spectrum, the semiclassical analysis is
not applicable. One has to reply therefore on a complementary
method which we will employ below. It will be applicable for the
spectral parameter $u \sim \mathcal{O} (\eta^0)$ thus yielding a
valid expression for the anomalous dimensions which can be used
for both the lower and upper parts of the spectrum.

Our subsequent consideration will be focused on developing different techniques
to analyze the excited states in the vicinity of the lower and upper boundary
of the band. The consideration will be limited to the lowest three orders of
perturbation theory, though it can be generalized in a straightforward fashion
to even higher loops. At the lower boundary, the analysis will be performed in
great generality for any $N$ and $L$, while for the upper part we will consider
the three-site spin chain which, on the one hand, is simple and, on the other,
demonstrates all of the salient features of the method and emerging integrable
structures in higher orders of pertubration theory.

\section{Lower part of the spectrum}
\label{LowerPart}

Let us begin with the description of the lower part of the spectrum in the limit of large spin $N$.
The asymptotic solution to the Baxter equation in this region will be valid for the spectral
parameter behaving as $u \sim \mathcal{O} (\eta^0)$. The method relies on an observation that two
terms in the left-hand side of the Baxter equation \re{AllOrderBaxterEq} have different scaling
behavior as $\eta \to 0$. This method has been developed in Ref.~\cite{DerKorMan00} for short-range
spin chains and applied for the calculation of the one-loop minimal trajectory in
Ref.~\cite{BelGorKor06}. Our main goal is to extend this approach to higher loops. Let us
recapitulate the essential features of this formalism at one loop and then extend it to arbitrary
orders of perturbation theory.

\subsection{Asymptotic solution of one-loop Baxter equation}

We notice that the evaluation of the leading order anomalous dimension \re{LOgammaANDtheta} requires
the knowledge of the Baxter polynomial and its first derivative at the points $u = \pm \ft12 i$. In
their vicinity, due to the power suppression by the dressing factors, one can neglect either the
first or second term of the one-loop Baxter equation. This however persists in a more general
situation when $u \sim \mathcal{O} (\eta^0)$. In the asymptotic regime under consideration, the
transfer matrix is large $|t (u)| \gg 1$ since the conformal Casimir scales as a second power of
$\eta^{-1}$. Then, introducing the ratio of the Baxter polynomials $\varphi (u) = Q_0 (u + i)/Q_0
(u)$, one realizes that in order to match the scaling behavior of the left-hand side it should be
either large or small~%
\footnote{For the spectral parameter scaling as $u \sim \mathcal{O}
(\eta^{- 1})$ both terms in the left-hand side would contribute on equal
footing.}. Due to this fact the second-order finite difference Baxter
equation \re{LoBaxterEq} at one-loop level splits into two first-order
finite-different equations \cite{BelGorKor06}
\be
\label{LOBax+-} (u^\pm)^{L} Q_0^\pm (u \pm i) = t_0 (u) Q_0^\pm (u) \, ,
\ee
whose solutions have the form (up to an overall normalization factor),
\be
\label{LOsols}
Q_0^\pm (u)
=
2^{\mp iu}
\frac{
\prod_{k = 1}^L\Gamma \left( \mp i u \pm i \delta_{k, 0} \right)}
{\Gamma \left(\mp i u + \ft12 \right)^L}
\, ,
\ee
where we used the parameterization of the transfer matrix at leading
order in terms of its roots $\delta_{k, 0}$,
\be
\label{trmatr}
t_0 (u)
=
2 \prod_{k=1}^{L} (u - \delta_{k, 0})
\, .
\ee
The asymptotic solution of the one-loop Baxter equation is given by a linear combination of $Q_0^\pm
(u)$,
\begin{eqnarray}
\label{as-sol-LO}
Q_0^{\rm (as)} (u) = A^+_0 Q_0^+ (u) + A^-_0 Q_0^-(u)
\, ,
\end{eqnarray}
where $A^\pm_0$ are fixed up to an overall normalization from the condition of
the cyclic symmetry of the eigenstates
\be\label{cycl}
\exp(i\theta_0) =\frac{Q_0 (+ \ft{i}{2})}{Q_0 (- \ft{i}{2})}
= 1 \quad \Longrightarrow \quad A^\pm_0= Q_0^{\mp}(\mp \ft12 i)
\, .
\ee

\subsection{Asymptotic solution of all-loop Baxter equation}

The asymptotic solution to the one-loop Baxter equation can be generalized to
higher loops. To this end we have to assume that higher loop corrections to
the conserved charges $q_k (g)$, roots of transfer matrix $\delta_k (g)$ and
Bethe roots $u_k(g)$ are compatible with the large-$N$ scaling of the
corresponding one-loop quantities, i.e., higher loop corrections should not
violate their scaling behavior at one loop so that $|t (u)| \gg 1$
order-by-order in 't Hooft coupling. Analogously to the leading order
equation, expanding both sides of the Baxter equation \re{AllOrderBaxterEq}
in powers of 't Hooft coupling, one observes that in the vicinity of the
points $u = \pm \ft12 i$ one can neglect either set of terms arising from
the expansion of the two terms in the left-hand side of the all-loop Baxter
equation \re{AllOrderBaxterEq}. A close inspection shows that, contrary to
the one-loop case where it is valid for any length $L$ of the chain, this
assumption is fulfilled starting from certain values of $L$ which is larger
at higher orders of perturbation theory.

Having these limitations in mind, the asymptotic solution to the all-loop
Baxter equation \re{AllOrderBaxterEq} can be split into two equations
\be\label{EQ1}
(x^\pm)^L {\rm e}^{\Delta_\pm (u^\pm)} Q^\pm (u \pm i) = t (u) Q^\pm (u)
\, .
\ee
Now we proceed with the solution of these equations in the first three orders
of the loop expansion. Taking into account that the only dependence of the
Baxter polynomial and the transfer matrix on the coupling constant comes
through their roots, the perturbative expansion of the latter reads
\be
Q (u) = \sum_{n = 0} g^{2 n} Q_n (u)
\, , \qquad
t (u) = \sum_{n = 0} g^{2 n} t_n (u)
\, ,
\ee
where the leading order transfer matrix $t_0 (u)$ is given by Eq.\ \re{trmatr}
while the higher terms are
\ba
t_1 (u)
\!\!\!&=&\!\!\!
- \left( \sum_{j = 1}^L \frac{\delta_{j,1}}{u - \delta_{j,0}} \right) t_0 (u)
\, , \\
t_2 (u) \!\!\!&=&\!\!\! - \left( \sum_{j = 1}^L
\frac{\delta_{j,2}}{u - \delta_{j,0}} - \sum_{j < n}^L
\frac{\delta_{j,1}}{u - \delta_{j,0}} \frac{\delta_{n,1}}{u -
\delta_{n,0}} \right) t_0 (u) \, , \qquad \dots \, . \nonumber
 \ea
Substituting these relations into the asymptotic Baxter
equations \re{EQ1} and expanding the dressing factors to the required order,
we deduce the first-order finite difference equations for $Q_1 (u)$
\ba \label{EQ2}
(u^\pm)^{L} Q_1^\pm (u \pm i) \!\!\!&-&\!\!\!
(u^\pm)^{L - 2} \left( \ft{1}{4} L \mp i u^\pm \gamma_0^\pm \right)
Q_0 (u \pm i) = Q_1^\pm (u) t_0 (u) + Q_0^\pm (u) t_1 (u)
\, ,
\ea
and $Q_2 (u)$
\begin{align}
\label{EQ3}
 & (u^\pm)^{L} Q_2^\pm (u \pm i) -(u^\pm)^{L - 2} \left( \ft{1}{4} L \mp i u^\pm
\gamma_0^\pm \right) Q_1^\pm (u \pm i)
\\
 + &(u^\pm)^{L - 4} \left( \ft{1}{32} L (L - 3) \mp \ft{i}4 (L - 1) u^\pm \gamma_0^\pm \mp i
(u^\pm)^2 \alpha_1^\pm \pm i \gamma_1^\pm (u^\pm)^3 \right) Q_0^\pm (u \pm i)
\nonumber\\
& \hspace*{56mm} = Q^\pm_2 (u) t_0 (u) + Q^\pm_1 (u) t_1 (u) + Q^\pm_0 (u) t_2 (u) \, . \nonumber
\end{align}
In these equations $\gamma^\pm_\ell $ stand for the coefficients in the
perturbative expansion of the anomalous dimensions \re{AllOrderAD}
\be
\gamma (g) = \sum_{n = 0} g^{2 n+2} \gamma_{n} \,,\qquad  \gamma_n = \gamma^+_n - \gamma^-_n \, ,
\ee
with $\gamma^\pm_\ell $ associated with $\Delta^{(1)}_\pm (g) $ in
\re{AllOrderAD}, and the  constants $\alpha_1^\pm$ are defined as
 \be
\alpha_1^\pm = \pm\frac{3 i}{16} \left( \frac{Q_0^\prime
(\pm\ft{i}{2})}{Q_0 (\pm\ft{i}{2})} \right)^2 \mp \frac{i}{16}
\frac{Q_0^{\prime\prime} (\pm\ft{i}{2})}{Q_0 (\pm\ft{i}{2})} \, .
\ee
As can be seen from the structure of the asymptotic Baxter equations \re{EQ2}
and \re{EQ3}, beyond one loop level  it is natural to look for the solution
for $Q^\pm_\ell (u)$ in the following factorized form
\begin{eqnarray}
\label{QG} Q^\pm_n (u) = Q^\pm_0 (u) G^\pm_n (u) \, , \qquad n > 0 \, .
\end{eqnarray}
The solutions are easily constructed in term of Euler psi-functions and read,
\ba \label{NLOsols} G_1^{\pm} (u) \!\!\!&=&\!\!\!
c_1^{\pm} + \ft14 L \psi^\prime (\mp i u + \ft12) - \gamma_0^{\pm}
\psi (\mp i u + \ft12) \pm i \sum_{k = 1}^L \delta_{k, 1} \psi (\mp
i u \pm i \delta_{k, 0})
\, , \\ \notag
G_2^{\pm} (u) \!\!\!&=&\!\!\! c_2^{\pm} + \ft12 \left( G_1^{\pm} (u)
\right)^2 - \ft12 \sum_{j = 1}^L \delta_{j,1}^2 \psi^\prime (\mp i u
\pm i \delta_{j,0} ) \pm i \sum_{j = 1}^L \delta_{j,2} \psi (\mp i u
\pm i \delta_{j,0} )
\\
&-&\!\!\! \ft{1}{64} L \psi^{\prime\prime\prime} (\mp i u + \ft12 )
+ \ft{1}{8} \gamma_0^{\pm} \psi^{\prime\prime} (\mp i u + \ft12 ) +
\left( \pm i \alpha_1^{\pm} - \ft12 ( \gamma_0^{\pm} )^2 \right)
\psi^{\prime} (\mp i u + \ft12 ) - \gamma_1^{\pm} \psi (\mp i u +
\ft12 ) \, . \nonumber
\ea
Note that the two and three-loop asymptotic functions $G_{1,2}^{\pm}(u)$ contain
arbitrary constants $c^{\pm}_{1,2}$ reflecting the fact that the solutions of
the asymptotic Baxter equation \re{EQ1} are defined modulo multiplication by an
arbitrary constant $ Q^\pm (u)\to c^\pm(g)  Q^\pm (u)$ with $c^\pm(g)  = 1 +
c_1^\pm g^2 + c_2^\pm g^4 +\ldots$. As we will see momentarily, although the
constants $c^{\pm}_{1,2}$ appear in the solutions, they do not contribute to
the anomalous dimensions.

The asymptotic solution to the Baxter equation is given by a linear combinations
of $Q^\pm (u)$
\be
Q^{\rm (as)} (u) = A^+ (g)
Q^+ (u) + A^- (g) Q^- (u) \, ,
\ee
with the coefficients $A^\pm (g)$ admitting perturbative expansion in the coupling
constant
\be
A^\pm
(u) = A^\pm_0 (u) + g^2 A^\pm_1 (u) + g^4 A^\pm_2 (u) + \mathcal{O}
(g^6) \, ,
\ee
and with analogous expansions for $Q^{\rm (as)} (u)$
\ba
Q_0 (u) \!\!\!&=&\!\!\! A^+_0 Q^+_0 (u) + A^-_0 Q^-_0 (u)
\, , \\ \notag
Q_1 (u) \!\!\!&=&\!\!\! A^+_1 Q^+_0 (u) + A^-_1 Q^-_0 (u) + A^+_0 Q^+_1 (u) +
A^-_0 Q^-_1 (u)
\, , \quad \dots \, .
\ea
Recall that to one loop, the normalization coefficients $A_0^\pm$ are fixed by
the cyclicity condition $\exp(i \theta) = 1$, Eq.~\re{cycl}. In a similar
manner, constraints on the normalization parameters $A^\pm_{1,2}$ come from
the fact that the quasimomentum $\theta$ is protected from perturbative
corrections and, therefore, it does not depend on the 't Hooft coupling
constant $g$.

\subsection{Asymptotic anomalous dimensions}

Using the expression for the anomalous dimensions in terms of the
all-order Baxter polynomial \re{AllOrderAD}, which explicitly reads
\be
 \gamma (g) = i g^2 \int_{- 1}^1 \frac{d t}{\pi} \sqrt{1 - t^2}
\left( \ln \frac{Q ( + \ft{i}{2} - g t )}{ Q ( - \ft{i}{2} - g t )}
\right)^\prime =  \gamma^+ (g) -\gamma^-(g) = 2 \Re{\rm e}\,  \gamma^+ (g) \, ,
\ee
one immediately finds the first three terms in 't Hooft expansion
\ba \label{NLO-NNLO-en}
\gamma_0^\pm
\!\!\!&=&\!\!\! \ft{i}2\left( \ln Q_0 (\pm \ft{i}{2}) \right)^\prime
\, , \\ \notag
\gamma_1^\pm \!\!\!&=&\!\!\! \ft{i}4\left( G_1^{\pm} (\pm \ft{i}{2})
\right)^\prime + \ft{i}{16} \left(\ln Q^{\pm}_0 (\pm \ft{i}{2})
\right)^{\prime\prime\prime}
\, , \\
\gamma_2^\pm \!\!\!&=&\!\!\! \ft{i}8 \left( G_2^{\pm} (\pm \ft{i}{2})
\right)^\prime - \ft{i}8 G_1^{\pm} (\pm \ft{i}{2}) \left( G_1^{\pm}
(\pm \ft{i}{2}) \right)^\prime + \ft{i}{32} \left( G^{\pm}_1 (\pm
\ft{i}{2}) \right)^{\prime\prime\prime} + \ft{i}{384} \left( \ln
Q^{\pm}_0 (\pm \ft{i}{2}) \right)^{(5)} \, .
\nonumber
\ea
Substituting the asymptotic solutions \re{LOsols} and \re{NLOsols} into \re{NLO-NNLO-en} we
immediately arrive at the following expressions for the anomalous dimension in terms of the roots of
the transfer matrix $\delta_k$
\ba
\label{LOAD}
\gamma_0 \!\!\!&=&\!\!\! \ln 2 - L \psi(1) +
\Re{\rm e} \sum_{k = 1}^L \psi ( \ft12 + i \delta_{k, 0} )
\, , \\
\label{NLOAD}
\gamma_1 \!\!\!&=&\!\!\! \ft{3}{8} L
\psi^{\prime\prime} (1) - \ft12 \gamma_0 \psi^\prime (1) - \ft{1}{8}
\Re{\rm e} \sum_{k = 1}^L \psi^{\prime\prime} ( \ft12 + i
\delta_{k,0} ) - \, \Im{\rm m} \sum_{k = 1}^L \delta_{k, 1}
\psi^\prime ( \ft12 + i \delta_{k, 0} )
\, , \\
\label{NNLOAD}
\gamma_2 \!\!\!&=&\!\!\!
\ft{1}{8} \psi^{\prime\prime\prime} (1) \gamma_0
-
\ft12 \psi^{\prime} (1) \gamma_1
-
\ft12 \Re{\rm e} \left( (\gamma_0^+)^2
-
2 i \alpha_1^+ \right) \psi^{\prime\prime} (1)
-
\ft5{96} L \psi^{(4)} (1)
+
\ft{1}{192} \Re{\rm e} \sum_{k = 1}^L \psi^{(4)} ( \ft12 + i \delta_{k, 0} )
\nonumber\\
&+&\!\!\!
\ft{1}{8}
\Im{\rm m} \sum_{k = 1}^L \delta_{k, 1} \psi^{\prime\prime\prime} ( \ft12 + i
\delta_{k, 0} )
-
\ft12
\Re{\rm e} \sum_{k = 1}^L \delta_{k, 1}^2 \psi^{\prime\prime} ( \ft12 + i
\delta_{k, 0} )
-
\Im{\rm m} \sum_{k = 1}^L \delta_{k, 2} \psi^\prime ( \ft12 + i \delta_{k, 0} )
\, .
\ea
We see that though the solutions to the Baxter equation are plagued by the arbitrary constants
$c_{1,2}^{\pm}$ they disappear from the expressions for the anomalous dimensions \re{LOAD} --
\re{NNLOAD}. For the lowest trajectory in the spectrum with $q_3 = 0$, we immediately reproduce
findings of Ref.\ \cite{BecCat07}.

\subsection{Quantization conditions}

The anomalous dimensions \re{LOAD} -- \re{NNLOAD} depend on the roots of
the transfer matrix. In this section we will find their quantized values
which yield in turn the quantized charges $q_k$ parameterized by $L-2$
quantum numbers $\ell = \{ \ell_1, \ell_2, \dots , \ell_{L-2} \}$. Each
set of the charges gives rise to a certain state in the spectrum of
anomalous dimensions at fixed spin $N$ and twist $L$.

In the method of the Baxter $Q$-operator the quantization conditions follow from the requirement for
$Q (u)$ to be a polynomial in $u$. In the large-$N$ limit, or equivalently for $\eta\ll 1$, the
Bethe roots scale as $u_k \sim N \gg 1$ and therefore we are not allowed to impose the polynomiality
condition on $Q^{({\rm as})} (u)$ since this solution is only valid for the spectral parameter $u
\sim \mathcal{O} (\eta^0)$. Instead, we have to demand that the asymptotic solution $Q^{({\rm as})}
(u)$ has to be a regular function on the real axis for $u \sim \mathcal{O} (\eta^0)$. As it follows
from \re{QG} and \re{NLOsols}, possible poles in $Q^{({\rm as})} (u)$ originate only from the
$\psi$-functions on the right-hand side of \re{NLOsols} and their position is determined by roots of
the one-loop transfer matrix $\delta_{k,0}$. In the lower part of the spectrum, all but two roots
scale as $\delta_{k,0} \sim \mathcal{O} (\eta^0)$ (with $k=2,\ldots,L-1$) while the two remaining
(`large') roots scale as $\delta_{1,0}\sim -\delta_{L,0}\sim \mathcal{O} (\eta^{-1})$. We require
that $Q^{\rm (as)}(u)$ should have zero residues at `small' roots $u = \delta_{k,0} \sim \mathcal{O}
(\eta^0)$, $k = 2, \dots, L - 1$ and obtain the following quantization conditions
\begin{eqnarray}
\label{QC}
\res\limits_{u = \delta_{k,0}}
\left\{ A^+ (g) Q^+ (u) + A^- (g) Q^- (u) \right\}
=
0 \, ,
\end{eqnarray}
where the constants $A^\pm (g)$ are determined from the cyclicity
condition on the single-trace Wilson operators.

At one-loop level, we take into account the relations \re{LOsols} and
\re{cycl} and obtain from \re{QC} the quantization conditions on the
roots of the one-loop transfer matrix
\cite{BelGorKor06}
\be
\label{quant_cond_LO}
{\rm e}^{- 2 i \delta_{j,0}}
\prod_{k \neq j}^L
\frac{
\Gamma (i \delta_{k,0} - i \delta_{j,0})
}{
\Gamma (i \delta_{j,0} - i \delta_{k,0})
}
\left[
\frac{\Gamma ( \ft12 + i \delta_{j,0})
}{
\Gamma ( \ft12 - i \delta_{j,0})
}
\right]^L
\prod_{k = 1}^L
\frac{
\Gamma (\ft12 - i \delta_{k,0})
}{
\Gamma (\ft12 + i \delta_{k,0})
}
=
1
\, .
\ee
Note that the product over $j$ involves all roots, large and small, while the
conditions are imposed on the small roots only. For the minimal energy we
have only two large roots $\delta_{1,0} = - \delta_{L,0}$, while the rest
of the roots are small and paired $\delta_{k,0} = - \delta_{L-k+1, 0}$ with
$k = 2, \dots, L - 1$. Separating the contribution of the large roots and taking
the logarithm of \re{quant_cond_LO} we get
\begin{eqnarray}
\label{QC-ln}
\delta_{n,0} \, \ln (-q_{2,0})
+ L\, \mathrm{arg}\,\Gamma(\ft12-i\delta_{n,0})+
\sum^{L-1}_{j=2}\,\mathrm{arg}\,\Gamma(1+i\delta_{n,0} - i \delta_{j,0})=
\ft12 \pi k_n
\, ,
\end{eqnarray}
which can be considered as an equation for the small leading order roots
$\delta_{n,0}$. The integer numbers $k_2 > k_3 > \dots$ count the
branches of the logarithms and play the role of quantum numbers in the
system of $L-2$ quantized charges. They satisfy the condition $k_n =
- k_{L-n+1}$. The explicit dependence of integers $k_n$ on their
index $n$ for the minimal energy will be obtained in the next
subsection. Expanding Eq.\ \re{QC-ln} in the limit $|\delta_{n, 0}|
\ll 1$, we finally arrive at the following quantization condition
\cite{BelGorKor06}
\begin{eqnarray}
\label{LO-QC-fin}
\delta_{n, 0} \simeq
\frac{\ft12 \pi k_n}{\ln(-q_{2, 0}) + (L - 2) \psi(1) - L \psi(\ft12)}
\, .
\end{eqnarray}
The exact two- and three-loop quantization conditions are more involved.
To save space, we present here only the two-loop result
\ba
\Re{\rm e}
\bigg\{
\delta_{j,1} [ \ln 2
\!\!\!&-&\!\!\!
L \psi (\ft12 - i \delta_{j,0}) ]
-
\sum_{k = 1}^L
\left[
( \delta_{k,1} - \delta_{j,1}) \psi (i \delta_{j,0} - i \delta_{k,0})
-
\delta_{k,1} \psi (\ft12 - i \delta_{k,0})
\right]
\bigg\}
\\[-2mm]
&=&\!\!\! \Im{\rm m}
\bigg\{
\gamma^+_0 \left( \psi (1) - \psi (\ft12 - i \delta_{j,0}) \right)
+ \ft{1}4 L \psi^\prime (\ft12 - i \delta_{j,0})
- \ft{1}{4} \sum_{k = 1}^L \psi^\prime (\ft12 - i \delta_{k,0})
\bigg\}
\, . \nonumber
\ea
Similarly to Eq.\ \re{LO-QC-fin}, these quantization conditions simplify
for small higher order roots. Keeping only the leading terms in their
Taylor expansion, they read at two- and three-loop orders, respectively,
\begin{eqnarray}
\label{2LQC-fin}
\left[ 2 \ln( \sqrt{2}\delta_{1,0} ) - L \psi(\ft12) + (L-2) \psi(1)
\right]
\delta_{n,1} \!\!\!&=&\!\!\! \ft14 \left( 2\gamma_0
\psi^\prime (\ft12) - L \psi^{\prime\prime} (\ft12) \right)
\delta_{n, 0} + \mathcal{O}( \delta^2 )
\, , \nonumber\\
\label{3LQC-fin}
\left[ 2 \ln(\sqrt{2} \delta_{1,0}) - L \psi(\ft12)
+ (L - 2) \psi(1) \right]
\delta_{n,2} \!\!\!&=&\!\!\! \ft14 \left(
2\gamma_0 \psi^\prime (\ft12) - L \psi^{\prime\prime} (\ft12)
\right) \delta_{n,1}
\\
+ \ft{1}{64} \Big[ L \psi^{(4)} (\ft12) - 8 \psi^{\prime\prime}
(\ft12) \gamma_0^2 - 4 \psi^{\prime\prime\prime} (\ft12) \gamma_0
\!\!\!&+&\!\!\! 32 \gamma_1 \psi^\prime (\ft12) - 2
(\chi^{+}+\chi^{-}) \psi^{\prime\prime} (\ft12) \Big] \delta_{n, 0}
\nonumber\\
+ \ft{i}{16} (c_1^{+} \!\!\!&-&\!\!\! c_1^{-}) (2\gamma_0 \psi(1) -
L \psi^\prime (1) + 4 \pi \delta_{1,1} - c_1^{+} - c_1^-)
+
\mathcal{O}(\delta^2) , \!\! \nonumber
\end{eqnarray}
up to higher order terms in small roots, denoted cumulatively by $\mathcal{O}(\delta^2)$. These
linear equations relate small roots of the transfer matrix at successive orders of perturbation
theory with coefficients accompanying them which scale logarithmically with $N$ stemming from the
anomalous dimensions and logarithms of the large roots of the transfer matrix. To simplify
notations, we introduced the following combination
\begin{eqnarray}
\chi^{\pm}=16(\gamma^{\pm}_0)^2 \pm 16 i \alpha_1^{\pm} = \left(
\frac{Q_0^\prime (\pm\ft{i}{2})}{Q_0 (\pm\ft{i}{2})} \right)^2 +
\frac{Q_0^{\prime\prime} (\pm\ft{i}{2})}{Q_0 (\pm\ft{i}{2})} \, .
\end{eqnarray}

We notice that at large $N$ the coefficients in front of $\delta_{n,1}$ and $\delta_{n,0}$ in both
sides of the first relation in \re{2LQC-fin} are of the same order $\mathcal{O}(\ln N)$ and, therefore,
$\delta_{n,1}/\delta_{n,0}= O(N^0) $. We also
observe that the two-loop constants $c_1^{\pm}$ do not show up in the two-loop condition, however,
they enter into the three-loop quantization condition, the second relation in \re{3LQC-fin}. A
quick inspection shows that while all $c_1^\pm$-independent
terms are of order one%
\footnote{The apparent dependence on $\ln^2 N$ rather than being linear in $\ln N$ in the second
term in the right-hand side of Eq.\ \re{3LQC-fin} cancels between $\gamma_0^2$ and $\chi^+ + \chi^-$
contributions, see Eq.\ \re{RelationsGammaChi}.}, the contribution proportional to $c_1^{+} -
c_1^{-}$ scales as $\ln N$, i.e.,
\begin{eqnarray}
2\gamma_0 \, \psi(1) - \ft16 \pi^2 L + 4 \pi \delta_{1,1} - c_1^{-} -c_1^+
\sim
2(2 \psi(1) + \sqrt{2} \pi) \ln{N} \gg \delta_{n, 0}
\, .
\end{eqnarray}
We recall that one of the starting points of the asymptotic method was the assumption that the
large-$N$ scaling of higher loop corrections to the $\delta$-roots should be compatible with the
scaling of the leading order roots $\delta_{k,0}$. In order for the three-loop quantization
conditions to be consistent with this assumption, or equivalently the corrections $\delta_{n,2}$ to
be ``small'', we have to impose the following constraint on the two-loop constants
\begin{eqnarray}\label{b1-conds}
c_1^{+} = c_1^{-}
\, .
\end{eqnarray}
Only in this case our initial assumption validating the current
approach will be satisfied. Fixing the relation between $c_1^{\pm}$
in this way and using in the leading order
\be
\label{RelationsGammaChi} \gamma_0 \simeq 2 \ln N \, , \quad \gamma_1 \simeq - \ft16 \pi^2 \ln N \,
, \quad \chi^{+} + \chi^{-} \simeq - 4 \gamma_0^2 \, ,
\ee
we finally get the two- and three-loop corrections to $\delta_n$
\begin{eqnarray}
\label{23L-QC}
\delta_{n,1}
\simeq
\ft14 \pi^2 \, \delta_{n,0}
\, , \quad
\delta_{n,2}
\simeq - \ft1{48} \pi^4 \delta_{n, 0} \, .
\end{eqnarray}
Equations (\ref{LO-QC-fin}) and (\ref{23L-QC}) together with the asymptotic
formulae for the anomalous dimensions \re{LOAD} -- \re{NNLOAD} completely
determine the spectrum at its lower boundary up to three loops.

\subsection{Ground state and daughter trajectories}

The expressions for the anomalous dimensions \re{LOAD} -- \re{NNLOAD} in terms
of the roots of transfer matrices are valid for $N, L \gg 1$ at the bottom of
the spectrum \re{twistLband}. Below we will analyze in detail the asymptotic
limit $\xi \gg 1$ and find preasymptotic corrections in $1/\xi$ to the leading
logarithmic behavior \re{twistLband}. To derive the explicit dependence of the
anomalous dimensions on the hidden parameter $\xi$, we heavily rely on known
scaling properties of the roots of the transfer matrix as a function of the
conformal spin found in the previous section. We remind that the large roots
are related to the eigenvalues of the conserved charge $q_2 (g)$,
\be
\delta_1 (g) = - \delta_L (g) = \left( - \ft12 q_2 (g) \right)^{1/2} + \dots \,
,
\ee
where $q_2 (g)$ is defined in Eq.\ \re{q2-def} such that
\ba
\delta_{1,0} = \frac{N}{\sqrt{2}} + \dots
\, , \quad
\delta_{1,1} = \ln N + \dots
\, , \quad
\delta_{1,2} = - \ft{1}{12} \pi^2 \ln N + \dots \, .
\ea
On the other hand, the small roots of the transfer
matrix scale as $\delta_k \sim 1/\xi$. Selecting the large roots
contributions and expanding the rest in terms of $|\delta_{n,0}| \ll
1$, we get for the one-loop minimal anomalous dimension the
well-known result \cite{BelGorKor06}
\ba \label{LO}
\gamma_0 = 2 \ln N - 2 L\ln 2 + 7\zeta(3) \sum_{j=2}^{L-1} \left( \delta_{k,0} \right)^2 + \dots \,
,
\ea
and analogously, applying (\ref{23L-QC}) to the two- and three-loop
anomalous dimensions we find
\begin{eqnarray}
\label{NLO}
\gamma_1
\!\!\!&=&\!\!\! - \ft16 \pi^2 \ln N
-
\ft72\zeta(3) + L \Big(\zeta(3)+\ft16\pi^2\ln 2\Big)
-
\Big( \ft{93}{2}\zeta(5) - \ft{35}{12}\zeta(3)\pi^2 \Big)
\sum_{j = 2}^{L - 1} \left( \delta_{k,0} \right)^2 + \dots \, , \qquad\\
\label{NNLO}
\gamma_2 \!\!\!&=&\!\!\! \ft{11}{360} \pi^4 \ln N
+
\ft{31}{4}\zeta(5) + \ft16\zeta(3)\pi^2
- L \Big(\ft{21}{8}\zeta(5)+\ft{1}{24}\zeta(3)\pi^2+\ft{11}{360}\pi^4\ln 2\Big)
\\
&&\qquad\qquad\qquad\qquad
+
\Big(\ft{1905}{8}\zeta(7)-\ft{155}{8}\zeta(5)\pi^2-
\ft{73}{720}\zeta(3)\pi^4\Big)
\sum_{k=2}^{L-1}(\delta^{(0)}_k)^2 + \dots \, , \nonumber
\end{eqnarray}
where the small roots $\delta_{k,0}$ are determined by the quantization
conditions (\ref{LO-QC-fin}). The ellipses stand for subleading terms
with higher powers of $\delta$'s. The quantized values of the small roots
$\delta_{k, 0}$ depend on the integer numbers $k_n = - k_{n-L+1}$ defined
in Eq.\ \re{QC-ln}, with the ones corresponding to the minimal anomalous
dimension being $k_n=L+1-2n$, yielding
\begin{eqnarray}\label{sum}
\sum_{j = 2}^{L-1} \left( \delta_{j, 0} \right)^2
\simeq
\frac{\pi^2}{16 \ln^2 N} \sum_{n = 2}^{L - 1} k_n^2
\simeq
\frac{L^3\pi^2}{48\ln^2{N}}
\, .
\end{eqnarray}

The set of relations \re{LO} -- \re{sum} defines the three-loop minimal anomalous dimension for $\xi
\gg 1$ with the first subleading corrections taken into account. The present formalism allows us to
calculate further terms in inverse powers of $\xi$, by expanding the Eqs.\ \re{LOAD} -- \re{NNLOAD}
to higher orders in small roots $\delta$. We will not perform this calculation further. The
anomalous dimensions of excited, or daughter, trajectories are determined by the same expressions
however with another set of integer numbers $k_n$. The first daughter trajectory possesses the same
``occupation numbers'' as the minimal anomalous dimension except for $k_2 = - k_{L-1} = L-1$. This
state is separated from the minimal one by the following gaps
\begin{eqnarray}
\Delta\gamma_0 \!\!\!&=&\!\!\! \ft72\zeta(3) \frac{L \pi^2}{\ln^2 N}
\, , \\
\Delta\gamma_1 \!\!\!&=&\!\!\! - \Big( \ft{93}{4}\zeta(5) -
\ft{35}{24}\zeta(3)\pi^2 \Big) \frac{L\pi^2}{\ln^2 N}
\, , \nonumber\\
\label{D} \Delta\gamma_2 \!\!\!&=&\!\!\!
\Big(\ft{1905}{16}\zeta(7)-\ft{155}{16}\zeta(5)\pi^2-
\ft{73}{1440}\zeta(3)\pi^4\Big)
\frac{L\pi^2}{\ln^2 N} \, , \nonumber
\end{eqnarray}
at one, two and three loops, respectively.

Substituting \re{sum} into \re{LO}, \re{NLO} and \re{NNLO} we find that the subleading corrections
to the ground state trajectory develop a nontrivial dependence on a new scaling parameter $j =
\xi^{- 1}=L/\ln N$
\begin{eqnarray}
\gamma^{(0)}_{N,L} (g) = \left[ 2 \Gamma_{\rm cusp} (g) + \varepsilon (g, j) \right] \ln N + \dots
\, .
\end{eqnarray}
As was shown in Refs.\ \cite{BelGorKor06,FroTirTse06,AldMal07}, this scaling behavior holds both at
weak and at strong coupling. In the large-$\xi$ limit, which we addressed above, $\varepsilon$
admits a double Taylor expansion in $g^2$ and $j$,
\be
\varepsilon (g, j) = g^2
\varepsilon_0 (j) + g^4 \varepsilon_1 (j) + g^6 \varepsilon_2 (j) + \dots \, ,
\ee
with perturbative coefficients taking the form of the power series in $j$,
\ba
\varepsilon_0 (j)
\!\!\!&=&\!\!\! - 2 \ln(2) j + \ft{7}{48} \pi^2 \zeta(3) j^3 + \dots
\, , \\[3mm]
\varepsilon_1 (j) \!\!\!&=&\!\!\! \left[ \zeta(3) + \ft{1}{6}
\pi^2 \ln 2 \right] j + \left[ - \ft{31}{32} \pi^2 \zeta(5) +
\ft{35}{576} \pi^4 \zeta(3) \right] j^3 + \dots
\, , \nonumber\\[3mm]
\varepsilon_2 (j) \!\!\!&=&\!\!\! - \left[ \ft{21}{8} \zeta(5) +
\ft{1}{24} \pi^2 \zeta(3) + \ft{11}{360} \pi^4 \ln 2 \right] j
+ \left[ \ft{635}{128} \pi^2 \zeta(7) - \ft{155}{384} \pi^4
\zeta(5) - \ft{73}{34560} \pi^6 \zeta(3) \right] j^3 + \dots
\nonumber
\ea
shown to order $j^4$. This result agrees with the findings of Ref.\ \cite{FreRejSta07} (see also
Refs.\ \cite{BasKor08,FioGriRos08,BucFio08,Gro08}).

\section{Upper part of the spectrum}
\label{UpperPart}

In the previous section, we described the lower part of the spectrum close to the minimal
trajectory. Let us turn to the study of the loop effects in the upper part of the band
\re{twistLband}. As we discussed earlier, in the asymptotic regime in question all roots of the
one-loop transfer matrix are large and scale like $\delta_{k, 0} = \mathcal{O} (N)$ implying that
conserved charges behave as $q_{k,0} \sim \eta^{- k}$, such that semiclassical expansion is
legitimate \cite{BelGorKor06}. As in the previous section, we shall assume that the scaling of
higher loop corrections to the $\delta$-roots is compatible with the one of leading order roots.

\subsection{Semiclassical Baxter equation}

In the considered asymptotic regime it is instructive to rescale the spectral
parameter
\be
u = \eta^{- 1} \widehat{u}
\, ,
\ee
and introduce the Hamilton-Jacobi ``action'' function $S (\widehat{u})$ as follows
\begin{eqnarray}\label{eikonal}
Q(u) = \exp \left( \frac{1}{\eta} S(\widehat{u}) \right)
\, ,
\end{eqnarray}
where
\be
S(\widehat{u}) = \eta \sum_{k = 1}^{N} \ln \big( \widehat{u} - \eta u_k (g)
\big)
\, .
\ee
One assumes that the rescaled charges $\widehat{q}_k (g) = \eta^k q_k (g)$
and the transfer matrix
\be
\label{TransferMatrixTau}
\tau(\widehat{u}) = \eta^L t(u)
=
2 \widehat{u}^L + \sum_{n = 1}^{L} \widehat{u}^{L - n} \widehat{q}_n (g)
\ee
admit a regular expansion in powers of $\eta$
\be
\label{q-tau-exps}
\widehat{q}_k (g)
=
\widehat{q}_k^{[0]} (g) + \eta \widehat{q}_k^{[1]} (g)
+
\mathcal{O} (\eta^2)
\, , \qquad
\tau (\widehat{u})
=
\tau^{[0]} (\widehat{u})
+
\eta \tau^{[1]} (\widehat{u})
+
\mathcal{O} (\eta^2)
\, ,
\ee
with each term having in turn a perturbative expansion in the coupling
constant. In the semiclassical approach one also assumes that the ``action"
$S(\widehat{u})$
is a series in $\eta$
\begin{eqnarray}\nonumber
S (\widehat{u})
=
S^{[0]} (\widehat{u}) + \eta S^{[1]} (\widehat{u})
+
\mathcal{O} (\eta^2)
\, ,
\end{eqnarray}
which is convergent and each term in it is uniformly bounded. As was
demonstrated in Ref.\ \cite{BelGorKor06} this assumption is justified
at leading order of the perturbative expansion anywhere in the band
provided that $\xi < 1$ or equivalently $\ln N < L$. In the case
$\xi \gg 1$, or $\ln N \gg L$ such an assumption breaks down in the
low-energy part of the spectrum and is valid therefore only at the
upper boundary. When going beyond one loop we assume similar behavior
for the power series in $\eta$ of higher order corrections in 't Hooft
coupling to the action $S^{[k]}_n (\widehat{u})$.

Substituting (\ref{eikonal}) and (\ref{q-tau-exps}) into the Baxter equation
\re{AllOrderBaxterEq} and expanding it in the double power series in $g^2$
and $\eta$ we get at leading order in 't Hooft coupling
\begin{eqnarray}
\label{WKB-LO}
2 \cos \left( S_0^{[0] \prime} (\widehat{u}) \right)
\!\!\!&=&\!\!\!
\frac{\tau_0^{[0]} (\widehat{u})}{ \widehat{u}^L}
\, , \\
\label{WKB-LO-eta}
- 2 S_0^{[1] \prime} (\widehat{u}) \sin \left( S_0^{[0] \prime} (\widehat{u})
\right)
\!\!\!&=&\!\!\!
\frac{\tau^{[1]}_0 (\widehat{u})}{\widehat{u}^L}
+
S_0^{[0] \prime\prime} (\widehat{u}) \cos \left( S_0^{[0] \prime} (\widehat{u})
\right)
+
\frac{L}{\widehat{u}} \sin \left( S_0^{[0] \prime} (\widehat{u}) \right)
\, .
\end{eqnarray}
Here $\tau_0^{[0]} (\widehat{u})/\widehat{u}^L$ plays the role of an effective
potential. Analogously, at two loops we obtain
\begin{eqnarray}
\label{WKB-NLO}
- 2 S_1^{[0] \prime} (\widehat{u}) \sin \left( S_0^{[0] \prime} (\widehat{u})
\right)
\!\!\!&=&\!\!\!
\frac{\tau_1^{[0]} (\widehat{u})}{ \widehat{u}^L}
\, , \\
- 2 S_1^{[1] \prime} (\widehat{u}) \sin \left( S_0^{[0] \prime} (\widehat{u})
\right)
\!\!\!&=&\!\!\!
\frac{\tau_1^{[1]} (\widehat{u})}{\widehat{u}^L}
-
S_1^{[0] \prime} (\widehat{u}) S_0^{[0] \prime\prime} (\widehat{u})
\sin \left( S_0^{[0] \prime} (\widehat{u}) \right)
\\
&+&\!\!\!
\left(
S_1^{[0] \prime\prime} (\widehat{u})
+
2 S_1^{[0] \prime} (\widehat{u}) S_0^{[1] \prime} (\widehat{u})
-
\frac{1}{\widehat{u}} \widehat{q}^{[0]}_{1,1}
+
\frac{L}{\widehat{u}}
S_1^{[0] \prime} (\widehat{u})
\right)
\cos \left( S_0^{[0] \prime} (\widehat{u}) \right)
\, . \nonumber
\end{eqnarray}
The derivation of subleading WKB corrections at higher loop orders is
straightforward but the resulting equations are rather cumbersome to be
displayed here.

\subsection{Bohr-Sommerfeld quantization}

The Bethe roots of the $SL(2)$ magnet take real values only. Numerical solution of the multiloop
Baxter equation suggests that this property persists in higher orders of perturbation theory and
eventually to all loops. In the semiclassical limit, $\eta \to 0$ the Bethe roots condense on finite
intervals on the real axis \cite{GauPas92,Kor95} so that their normalized distribution density
defined as
\begin{eqnarray}\label{density}
\rho (\widehat{u}) = \eta \sum_{k = 1}^N \delta \big( \widehat{u} - \eta u_k (g) \big) \, , \qquad
\int^{\infty}_{-\infty} d\widehat{u} \, \rho (\widehat{u}) = \eta N \, ,
\end{eqnarray}
vanishes outside of the domain
\begin{eqnarray}
\label{S-int} \mathcal{S} = [\widehat{\sigma}_1, \widehat{\sigma}_2] \cup
[\widehat{\sigma}_3,\widehat{\sigma}_4] \cup\dots\cup [\widehat{\sigma}_{2L -
3},\widehat{\sigma}_{2L - 2}] \, ,
\end{eqnarray}
with $\widehat{\sigma}_k$ obeying the condition $\widehat{\sigma}_k^{-L} \tau (\widehat{\sigma}_k) =
\pm 2$~\cite{Kor95}. Here the interval boundaries ordered as $\widehat{\sigma}_1 <
\widehat{\sigma}_2 < \dots < \widehat{\sigma}_{2L-2}$ and one of the intervals contains the origin.
The function $S'(\widehat{u})$ introduced in the previous section is related to the root density via
\begin{eqnarray*}
S'(\widehat{u}) = \int_{\mathcal{S}} d \widehat{v} \,
\frac{\rho (\widehat{u})}{\widehat{u} - \widehat{v}}
\, .
\end{eqnarray*}
It is a double-valued function on the complex $\widehat{u}$-plane with square-root branching points
$\widehat{\sigma}_k$.

Let us denote by $n_k$ the number of roots at the $k$-th interval $[\widehat{\sigma}_{2k-1},
\widehat{\sigma}_{2k}]$ with $k=1,\,...,\,L-1$. These integers do not depend on the coupling
constant simply because the number of roots at a given interval does not change when perturbative
corrections are turned on,
\begin{eqnarray}
\label{nCond}
\sum_{k = 1}^{L - 1} n_k = N
\, , \qquad
0 \leq n_1, \, n_2, \, \dots , \, n_{L-1} \leq N
\, .
\end{eqnarray}
By definition from (\ref{density}) and (\ref{S-int}) we have for each interval
from ${\cal S}$
\begin{eqnarray}
\label{rho-int}
\int_{\widehat{\sigma}_{2k-1}}^{\widehat{\sigma}_{2k}}
d \widehat{u} \, \rho(\widehat{u}) = \eta \, n_k
\, .
\end{eqnarray}
Expanding the density of roots in $\eta$-series $\rho (\widehat{u}) = \rho^{[0]}
(\widehat{u}) + \eta \rho^{[1]} (\widehat{u}) + \dots$ we obtain for small
values
of $n_k = \mathcal{O} (N^0)$ in each order of the expansion
\begin{eqnarray}
\int_{\widehat{\sigma}_{2k-1}}^{\widehat{\sigma}_{2k}}
d \widehat{u} \, \rho^{[0]} (\widehat{u}) = 0
\, , \quad
\int_{\widehat{\sigma}_{2k-1}}^{\widehat{\sigma}_{2k}}
d \widehat{u} \, \rho^{[1]} (\widehat{u}) = \eta \,n_k
\, , \quad
\int_{\widehat{\sigma}_{2k-1}}^{\widehat{\sigma}_{2k}}
d \widehat{u} \, \rho^{[j > 1]}(\widehat{u}) = 0
\, .
\end{eqnarray}
These can be rewritten as contour integrals of the resolvents over the cuts
$[\widehat{\sigma}_{2k-1}, \widehat{\sigma}_{2k}]$ on the real axis in the
complex $\widehat{u}$-plane,
\begin{eqnarray}
\label{rho-S-int}
\int_{\widehat{\sigma}_{2k-1}}^{\widehat{\sigma}_{2k}}
d \widehat{u} \, \rho (\widehat{u})
=
\oint_{\alpha_j} \frac{d \widehat{u}}{2 \pi i} S^\prime (\widehat{u})
\, ,
\end{eqnarray}
where the contours $\alpha_j$ encircle the intervals $[\widehat{\sigma}_{2k-1},
\widehat{\sigma}_{2k}]$ in the anticlockwise direction. As a result we obtain
the Bohr-Sommerfeld quantization conditions at all orders of perturbation
theory
\begin{eqnarray}
\label{QCforS}
\oint_{\alpha_j} \frac{d \widehat{u}}{2\pi i}  S^\prime (\widehat{u}) = \eta n_j
\, .
\end{eqnarray}
The set of integers $n = \{n_1, \dots, n_{L - 1}\}$ obeying \re{nCond} plays the
role of the quantum numbers determining the quantized values of the conserved
charges $q_k$ and, as a consequence, the anomalous dimensions of operators.
These numbers enumerate the trajectories which pass through anomalous dimensions
for integer values of the Lorentz spin $N$ and parameterize them starting from
the highest one residing at the upper boundary of the spectrum. These
trajectories represent a legitimate analytic continuation of the anomalous
dimensions. Another analytic continuation reflected in a yet another set of
trajectories encoded in the set of quantum numbers $\ell = \{ \ell_1, \ell_2,
\dots, \ell_{L - 2} \}$ was analyzed in the previous section. In fact the two
sets of integers are related to each other by linear relations, e.g., $\ell =
[N/2] - n$ for $L = 3$. Since the right-hand side of Eq.\ \re{QCforS} does not
depend on 't Hooft coupling $g$, we immediately conclude that
\begin{eqnarray}
\label{quant-cond-WKB-r}
\oint_{\alpha_j} \frac{d \widehat{u}}{2 \pi i}
S_0^\prime (\widehat{u})
=
\eta n_j
\, , \qquad
\oint_{\alpha_j} \frac{d \widehat{u}}{2 \pi i}
S_{n > 0}^\prime (\widehat{u})
= 0
\,
\end{eqnarray}
These relations hold to any order of the $\eta$-expansion and allow us to compute the quantized
charges $\widehat{q}_k^{[j]}$ for given $k$ and $j$. In the next subsection we will examine the two-
and three-loop quantization conditions (\ref{quant-cond-WKB-r}) order-by-order in powers of $\eta$.

\subsection{Quantization for three-site chain}

Let us resolve the quantization conditions for the high-spin Wilson operators of twist three, i.e.,
$L = 3$ and $N\ge 1$. In this case, the Bethe roots condense on two intervals. We assume that $n =
\mathcal{O} (N^0)$ roots reside on $[\widehat{\sigma}_1, \widehat{\sigma}_2]$ and the remaining $N -
n$ roots are on the interval $[\widehat{\sigma}_3, \widehat{\sigma}_4]$. Then in the leading order
of the semiclassical expansion, one finds that
\be
\oint_{\alpha_1} \frac{d \widehat{u}}{2 \pi i} S_0^{[0] \prime} (\widehat{u})
= 0
\, ,
\ee
which implies that the interval $[\widehat{\sigma}_1, \widehat{\sigma}_2]$
shrinks into a double point, according to the terminology of Section
\ref{AnalyticStructure}. The double point $\widehat{\sigma}^\ast$ is a
solution to the equations
\be
\tau^{[0] \prime}_0 (\widehat{\sigma}^\ast) = 0
\, , \qquad
| \tau_0^{[0]} (\widehat{\sigma}^\ast) | = 2
\, ,
\ee
and defines the critical values of the conserved charge
$\widehat{q}_{3,0}^{[0]}$
which takes the values
\be
\label{LOq3}
\widehat{q}^{[0]}_{3,0}=\pm\frac{1}{\sqrt{27}}
\, .
\ee
The two possible signs of the charge lead to the double degeneracy of the energy spectrum which is
not sensitive to the interchange $q_{3,0} \to - q_{3,0}$. We will see below that this degeneracy is
preserved in higher loops as well. For our subsequent consideration we choose therefore just one
value $\widehat{q}^{[0]}_{3,0} = 1/\sqrt{27}$.

Let us examine the Bohr-Sommerfeld quantization condition \re{quant-cond-WKB-r} involving the first
subleading semiclassical correction $S_0^{[1]} (\widehat{u})$ to the Hamilton-Jacobi function. Since
the integration contour $\alpha_1$ encircles a vanishing interval $[\widehat{\sigma}_1,
\widehat{\sigma}_2]$, the function $S_0^{[1]}$ should possess singularities on it to yield a nonzero
right-hand side in Eq.\ \re{quant-cond-WKB-r}. This can be demonstrated by opening up the double
point $\widehat{\sigma}^\ast = 1/\sqrt{12}$ into a small interval parameterized by a small parameter
$\varepsilon \ll 1$ and populated with $n$ Bethe roots
\be
[\widehat{\sigma}_1, \widehat{\sigma}_2] \, , \qquad \widehat{\sigma}_{1,2} =
\widehat{\sigma}^\ast
\mp \ft12 \varepsilon \, .
\ee
Using the condition defining the end-points of the cuts $\tau_0^{[0]}
(\widehat{\sigma}_{1,2}) = 2$, with the interval parameterized by the
variable $z$ such that $\widehat{u} = \widehat{\sigma}^\ast + \varepsilon z$,
we find in the leading order of the Taylor expansion in $\varepsilon$ that
\begin{eqnarray}
\widehat{q}^{[0]}_{3,0} \simeq \frac{1}{\sqrt{27}} - \frac{\sqrt{3}}{2} \,
\varepsilon^2
\, .
\end{eqnarray}
Substituting these results into Eqs.\ \re{WKB-LO} --
\re{WKB-LO-eta}, we can find the quantized values of the first
subleading corrections to $q_{3,0}$ from (\ref{quant-cond-WKB-r}),
\begin{eqnarray}
\varepsilon
\oint_{\alpha_1}
\frac{d z}{2\pi i} S^{[1] \prime}_0 (\widehat{\sigma}^\ast + \varepsilon z) = n
\, .
\end{eqnarray}
The contour $\alpha_1$ encircling the interval $[\widehat{\sigma}_1,
\widehat{\sigma}_2]$ can be deformed away from it, such that the integral
is given by the residue of $S^{[1] \prime}_0 (\widehat{\sigma}^\ast +
\varepsilon z)$ at $z = \infty$ keeping the product $\varepsilon z$ fixed.
Extending these considerations to higher orders in the $\eta$-expansion, we
get semiclassical corrections to the charge $\widehat{q}_{3}$ at one loop in
't Hooft coupling,
\begin{eqnarray}
\label{q3-ch}
\widehat{q}_{3,0}
\!\!\!&=&\!\!\!
\frac{1}{\sqrt{27}}
\\
&-&\!\!\!
\eta\frac{(n+1)}{\sqrt{3}}
+
\eta^2\frac{\sqrt{3}}{108}(24n^2+60n+43)
-
\eta^3\frac{\sqrt{3}}{324}(8n^3+48n^2+47n+32)
+ \dots\, . \nonumber
\end{eqnarray}
The procedure can be generalized to an arbitrary high order in $\eta$.

\begin{figure}[t]
\begin{center}
\mbox{
\begin{picture}(0,200)(110,0)
\put(0,0){\insertfig{7}{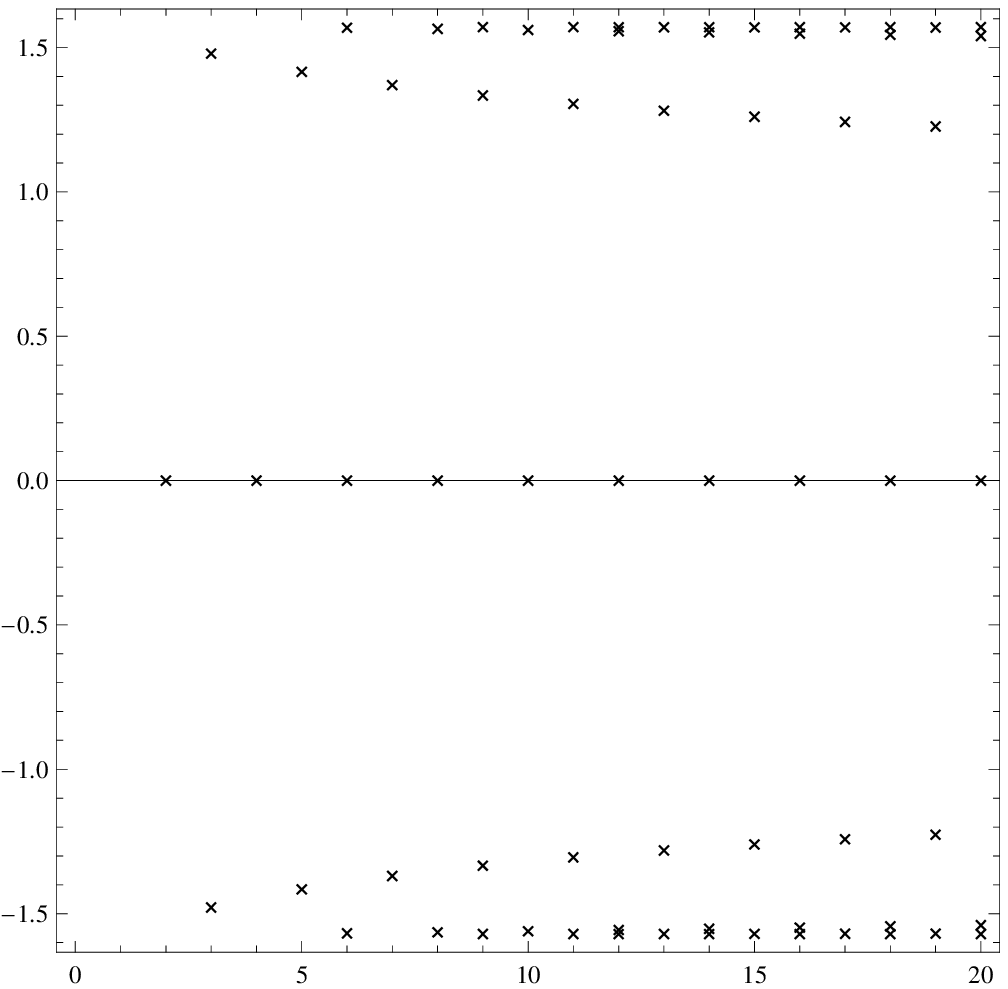}}
\put(-17,100){$\widehat{q}_{1,1}$}
\put(180,-10){$N$}
\end{picture}
}
\end{center}
\caption{\label{q1UPPER} The exact eigenvalues of charge $\widehat q_1$ at two-loop
order of perturbation theory.}
\end{figure}

Turning to higher orders of perturbation theory, one observes that starting from
two loops the transfer matrix \re{TransferMatrixTau} develops a nontrivial
coefficient in front of the $(L-1)$-st power of the spectral parameter, i.e.,
the charge $q_1 (g)$. The latter however is completely determined by the Baxter
polynomial \re{q1-def} such that the $n$-th perturbative term in $q_1$ is
defined by $(n-1)$-st order of $Q (u)$. Using the asymptotic solutions of the
leading and next-to leading order Baxter equations (\ref{LOsols}) and
\re{NLOsols}, it is straightforward to obtain $q_1 (g)$ up to three loops,
\begin{eqnarray}
\label{q1calc} \widehat{q}_{1,1} \!\!\!&=&\!\!\! \frac12i \eta
\sum_{j = 1}^L \left( \psi (\ft12 + i\delta_{j,0}) - \psi (\ft12 -
i\delta_{j,0}) \right)
\, , \\
\widehat{q}_{1,2}
\!\!\!&=&\!\!\!
- \frac{\pi^2}{12} \widehat{q}_{1,1}
\\
&+&\!\!\! \frac{i}{16} \eta\sum_{k=1}^L \left(
\psi^{\prime\prime} (\ft12 - i \delta_{k,0}) - \psi^{\prime\prime}
(\ft12 + i \delta_{k,0}) \right)
-
\frac14\eta \sum_{k=1}^L \delta_{k,1}
\left(
\psi^\prime (\ft12 - i\delta_{k,0}) + \psi^\prime (\ft12 + i \delta_{k,0})
\right) \, , \nonumber
\end{eqnarray}
where
\be
\delta_{k,0}
=
\frac{1}{\eta} \widehat{\delta}_{k,0}
=
\frac{1}{\eta}
\left(
\widehat{\delta}_{k,0}^{[0]}
+
\eta \widehat{\delta}_{k,0}^{[1]} + \mathcal{O}(\eta^2)
\right)
\, ,
\ee
are the leading order roots of the transfer matrix. They are exactly
calculable from the leading order charges $q_{k,0}$ for any fixed
$L$. Particularly, in the considered WKB regime one can explicitly show
from Eq.\ (\ref{q3-ch}) that all roots are ``large" and scale as
$\delta_{k,0} = \mathcal{O} (N)$. For the maximal energy and
$\widehat{q}_{3,0} > 0$, two of the roots are positive and one is
negative, $\widehat{\delta}_{1,0}^{[0]} > 0$,
$\widehat{\delta}_{2,0}^{[0]} > 0$, $\widehat{\delta}_{3,0}^{[0]} <
0$. Expanding the expression for two- and three-loop corrections
$\widehat{q}_{1, n}$ (\ref{q1calc}) in the limit $\eta\ll 1$, or
$|\widehat{\delta}_{k,0}^{[0]}/\eta|\gg 1$ and using the asymptotic
formula for the Euler digamma function
\begin{eqnarray*}
\psi(x)\simeq \ln{x}-\frac{1}{2}x^{-1}-\sum_{k=1}^{\infty}\frac{B_{2k}}{2k}{x^{-2k}} \, , \qquad x
\gg 1 \, ,
\end{eqnarray*}
we find (for $\hat{q}_{3,0} > 0$)
\begin{eqnarray}
\label{q1tot} \widehat{q}_{1,1} = - \frac12\pi \eta \, , \qquad
\widehat{q}_{1,2} = - \frac{\pi^2}{12} \widehat{q}_{1,1} \, .
\end{eqnarray}
Notice that $q_1$ changes its sign as $q_3 \to - q_3$ such that all explicit
expressions for $q_1$ we found above are accompanied by the signature factor
${\rm sign} (q_3)$ to accommodate both values of $q_1$. It is interesting to
note that $\widehat{q}_1$ does not receive any higher order power corrections
in $\eta$. Indeed, this result has been confirmed by numerical calculations
of the charge $\widehat{q}_{1,1}$ (see Fig.\ \ref{q1UPPER})

The resolution of the quantization conditions (\ref{quant-cond-WKB-r}) at higher
loops goes along the same lines as the one-loop calculation and yields the
quantized values of corrections to the conserved charge $q_3$ at every order
of the semiclassical expansion
\begin{eqnarray}
\label{q3tot}
\widehat{q}_{3,1}
\!\!\!&=&\!\!\!
\eta
\frac{\sqrt{3}}{36} \left( \sqrt{3} \pi + 6 \gamma^{[0]}_0 \right) -
\eta^2 \frac{1}{12} \left( \pi + 4\sqrt{3} \gamma^{[0]}_0 (n+1) -
2\sqrt{3} \gamma^{[1]}_0 - 6 \sqrt{3} \right)
\\
&+&\!\!\! \eta^3 \frac{\sqrt{3}}{432} \Big( 2\gamma^{[0]}_0 (24 n^2
+ 60 n + 43) - 144 \gamma^{[1]}_0 (n + 1) + 72 \gamma^{[2]}_0 + 216
n + 9 \sqrt{3} \pi \Big) + \dots
\, , \nonumber\\
\widehat{q}_{3,2} \!\!\!&=&\!\!\! \eta \frac{1}{24} \left( 4\sqrt{3}
\gamma^{[0]}_1 - \frac{\pi^3}{6} \right) - \eta^2
\frac{\sqrt{3}}{144} \Big( \pi^2 - 2\sqrt{3} \pi \left(
2\gamma^{[0]}_0 + \frac{\pi^2}{6} \right) + 48\gamma^{[0]}_1 (n + 1)
\nonumber\\
&&\qquad\qquad - 24 \, \gamma^{[1]}_1 - 12 \left( \gamma^{[0]}_0
\right)^2 - \frac32 \pi^2 \Big) + \dots \, . \nonumber
\end{eqnarray}
Here $\gamma_k^{[p]}$ is the $p$-th coefficient in the semiclassical expansion of the $k$-th order
anomalous dimension in 't Hooft coupling $g^2$,
\be
\gamma_k = \gamma_k^{[0]} + \eta \gamma_k^{[1]} + \eta^2 \gamma_k^{[2]} + \dots \, .
\ee
Several comments are in order.

\begin{figure}[t]
\begin{center}
\mbox{
\begin{picture}(0,140)(240,0)
\put(10,0){\insertfig{5}{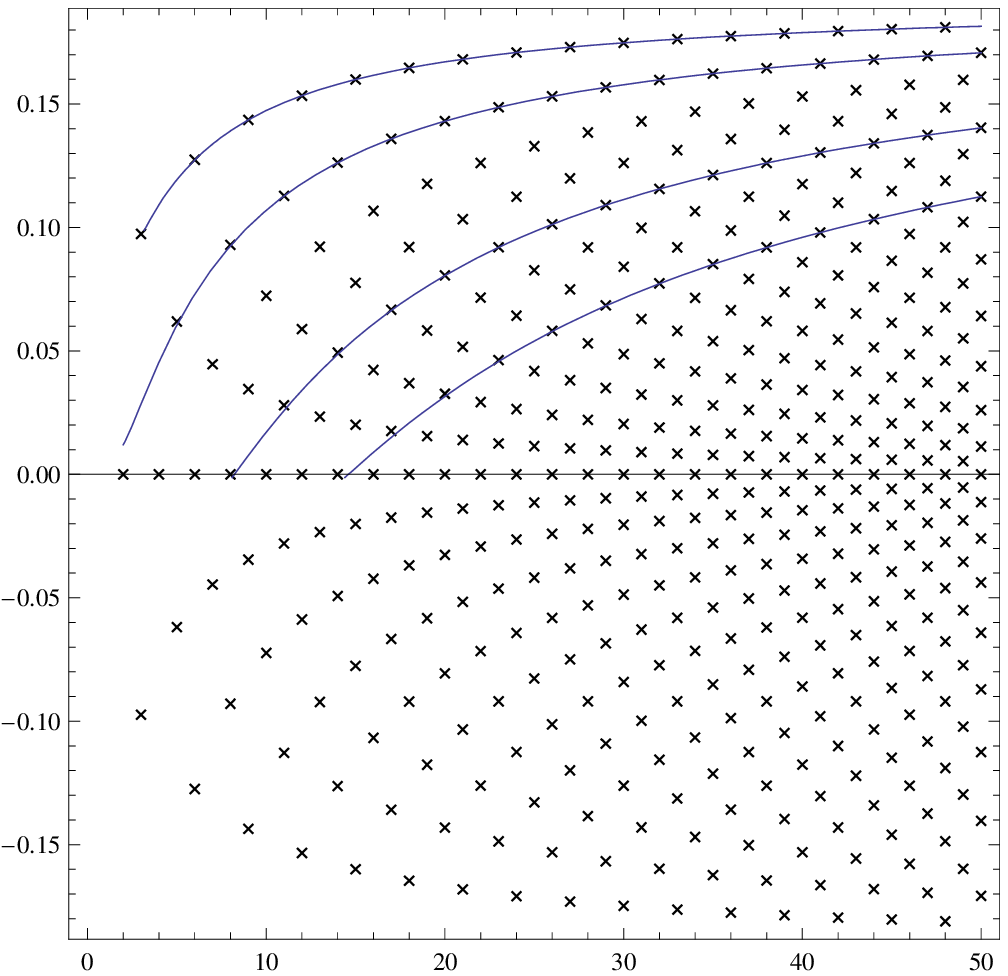}}
\put(-5,70){$\widehat{q}_{3,0}$}
\put(130,-10){$N$}
\put(175,0){\insertfig{5}{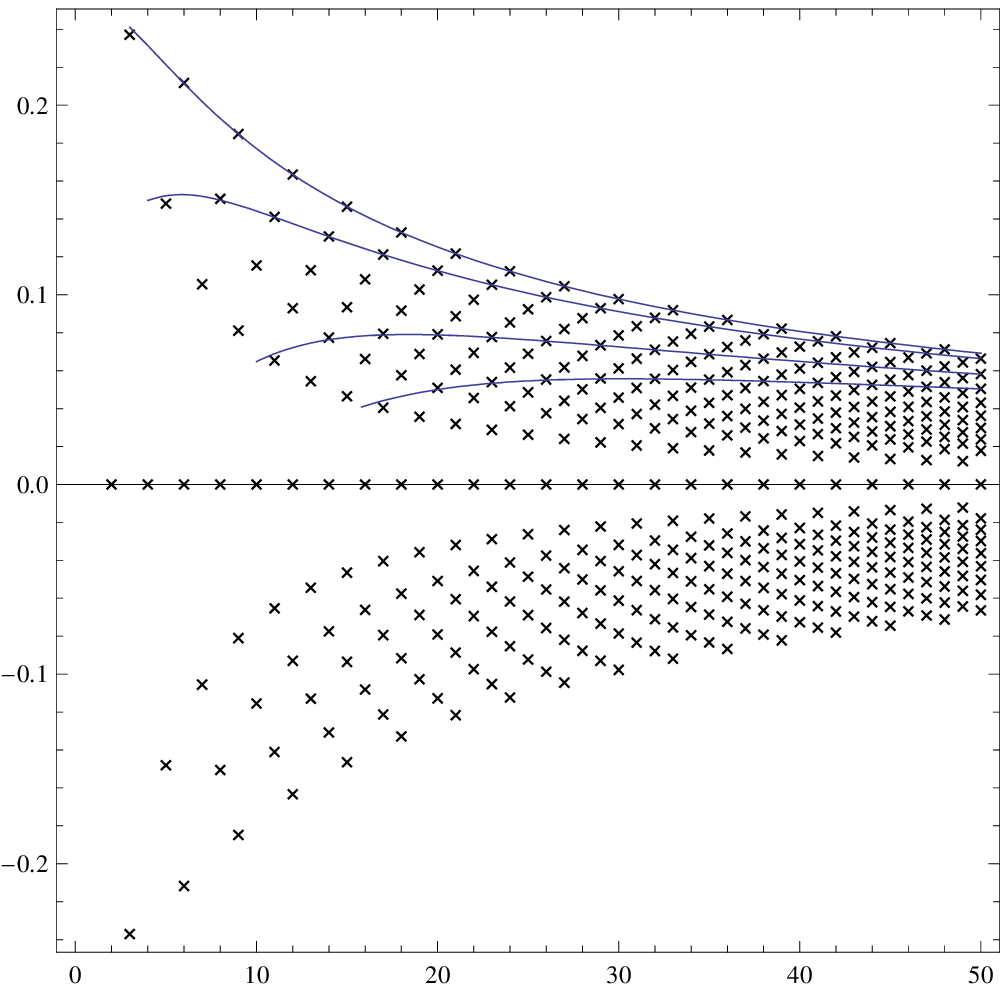}}
\put(160,70){$\widehat{q}_{3,1}$}
\put(295,-10){$N$}
\put(340,0){\insertfig{5}{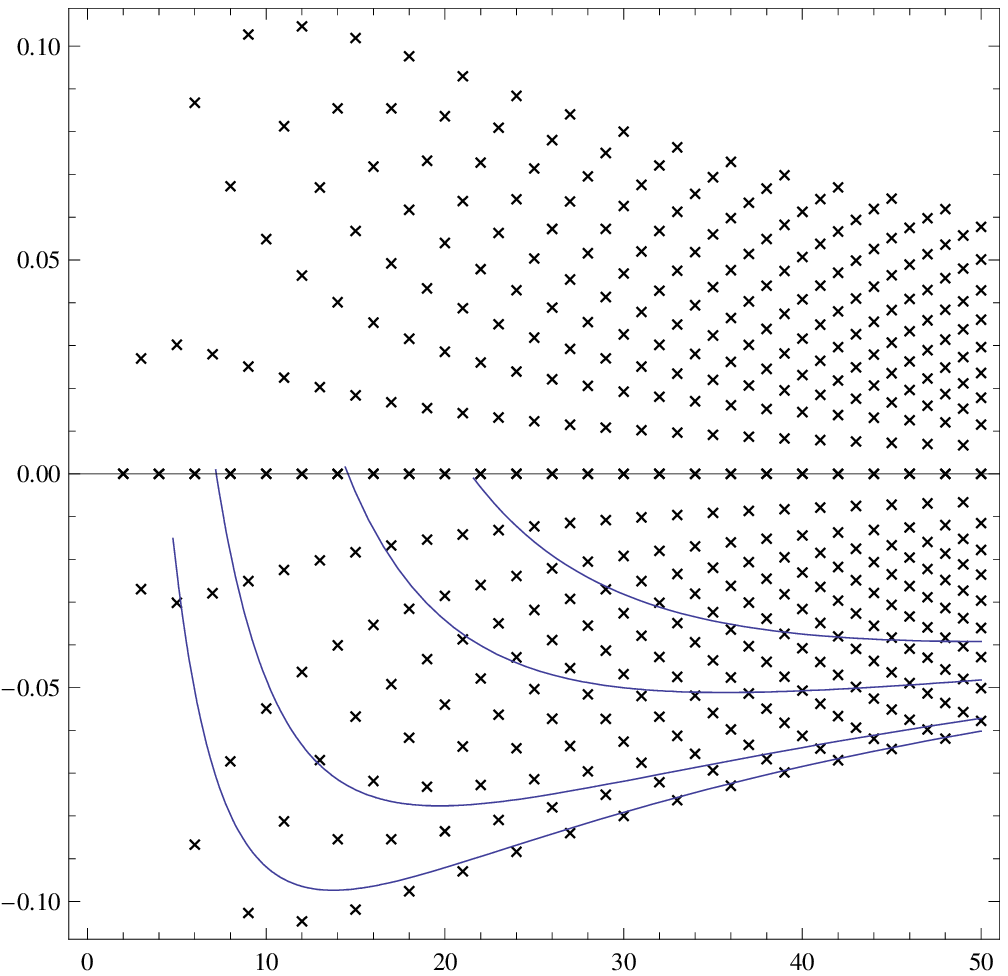}}
\put(325,70){$\widehat{q}_{3,2}$}
\put(460,-10){$N$}
\end{picture}
}
\end{center}
\caption{\label{NNLOq3s} The exact eigenvalues of the conserved charge $\widehat q_3$
at one-, two- and three-loop order (left to right) of perturbation theory for
twist-three operators shown together with a set of selected semiclassical
trajectories computed from Eqs.\ \re{q3-ch} and \re{q3tot}, respectively,
for $n=0,1,4,7$.}
\end{figure}

We observe that the semiclassical expansion of the two- and three-loop corrections to charges, i.e.
$\widehat{q}_{1, n}$, $\widehat{q}_{2, n}$ and $\widehat{q}_{3, n}$ (with $n\ge 1$), only starts
from the $\mathcal{O} (\eta)$ term. As anticipated, the fine structure of these charges, that is
their dependence on the integer $n$, resides in the subleading $\mathcal{O} (\eta^2)$ terms. The
semiclassical results \re{q3tot} provide a very accurate description of the exact spectrum of the
quantized charge $\widehat q_3$, see Fig.\ \ref{NNLOq3s}. For example, for $N=20$ we find the
following accuracy in each order of the $\eta$-expansion of $\widehat{q}_{3,1}$, Eq.~\re{q3tot},
\begin{eqnarray}\label{accur-R}
\eta: \quad + 4.77 \% \, , \qquad
\eta^2: \quad - 0.09\% \, , \qquad
\eta^3: \quad - 0.0003 \%
\, ,
\end{eqnarray}
confirming the fast converge of the semiclassical expansion.

\subsection{Semiclassical anomalous dimensions}

\begin{figure}[t]
\begin{center}
\mbox{
\begin{picture}(0,140)(240,0)
\put(10,0){\insertfig{5}{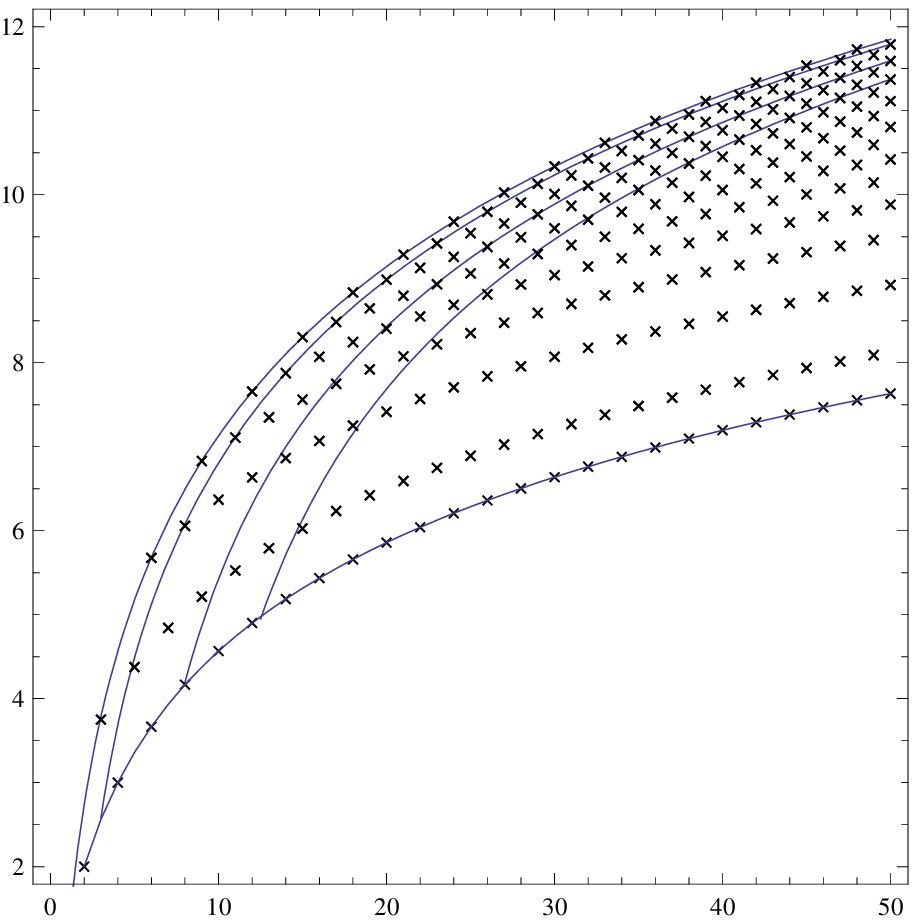}}
\put(-5,70){$\gamma_0$}
\put(130,-10){$N$}
\put(175,0){\insertfig{5}{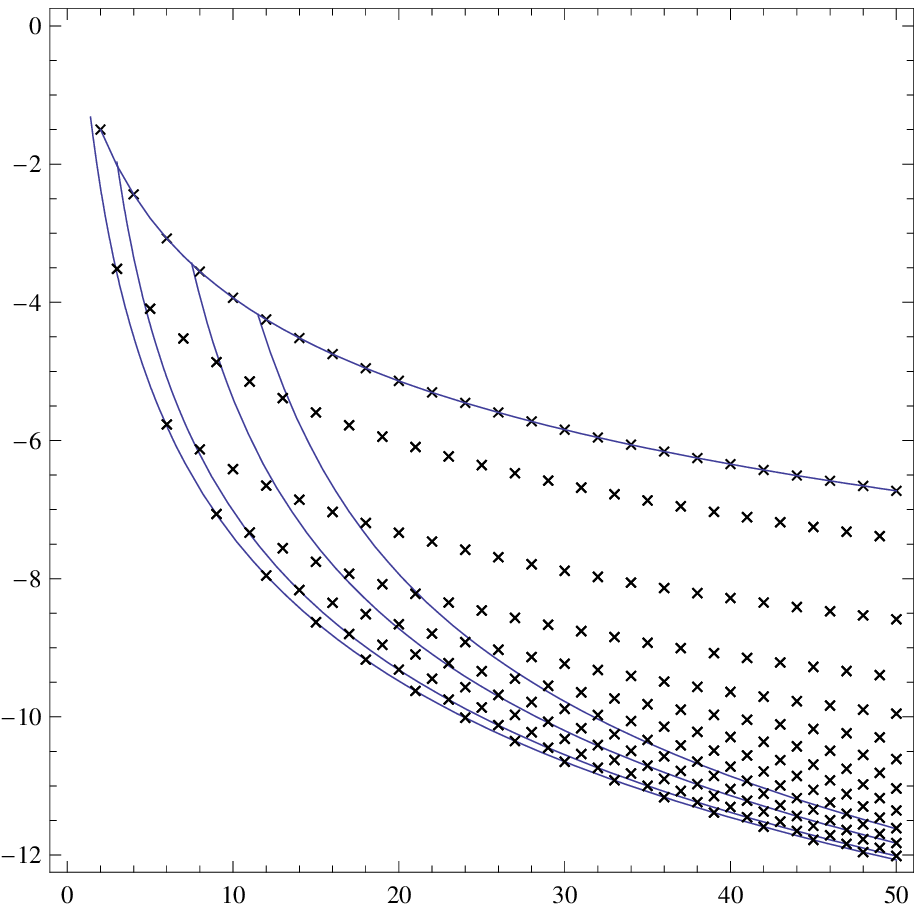}}
\put(160,70){$\gamma_1$}
\put(295,-10){$N$}
\put(340,0){\insertfig{5}{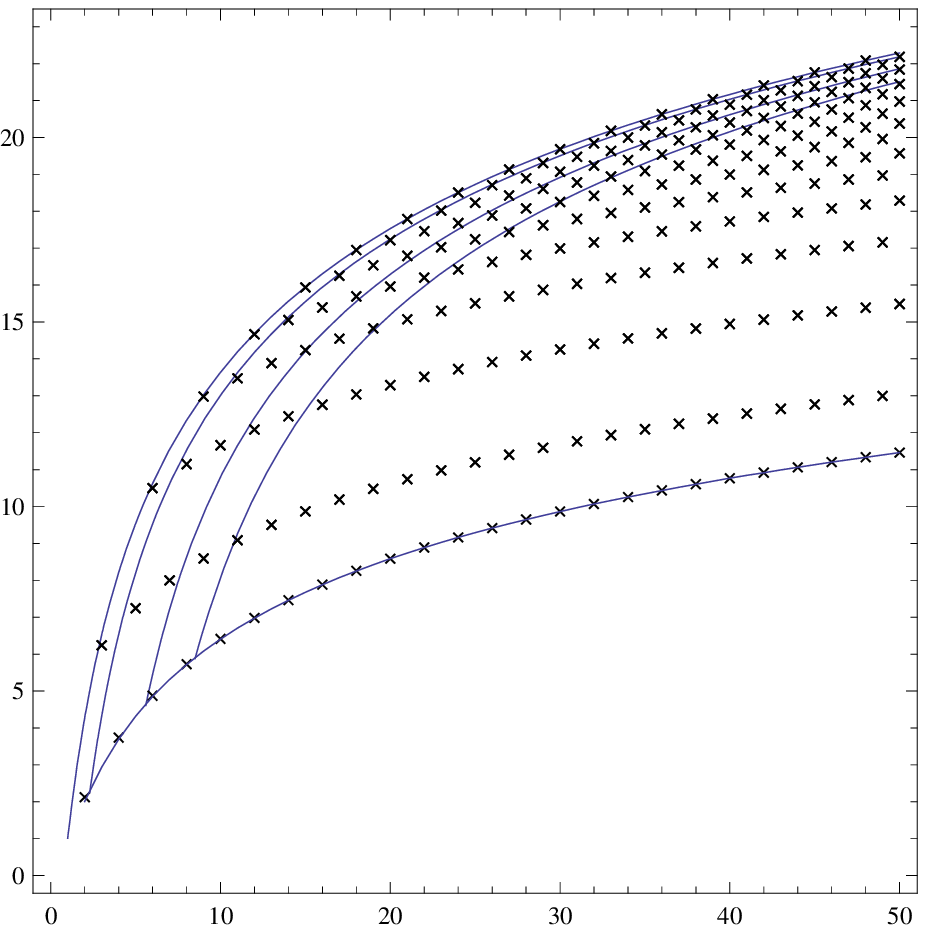}}
\put(325,70){$\gamma_2$}
\put(460,-10){$N$}
\end{picture}
}
\end{center}
\caption{\label{NNLOgammas} The exact eigenvalues of anomalous dimension matrix
in the lowest three orders of perturbation theory for twist-three operators
obtained from numerical solution of the Baxter equation and selected
semiclassical trajectories for $n=0,1,4,7$ at corresponding orders in 't Hooft
coupling determined by Eqs.\ \re{LOadUPPER} and \re{eps_fin}, respectively.}
\end{figure}

Having determined the quantized values of the conserved charges, we can immediately find
perturbative semiclassical expansion of the roots of the transfer matrix. When substituted into the
general expressions for the anomalous dimensions \re{LO} -- \re{NNLO}, it yields the trajectories in
the vicinity of the upper boundary of the spectrum enumerated by the integer $n$. Note that at any
loop level the argument of psi-functions contains only large leading order roots $\delta_{k,0}^{[0]}
= \mathcal{O}(N)$ so that one can find approximate formulas by expanding these $\psi$-functions in
the limit $\eta \ll 1$, or $|\widehat{\delta}_{k,0}^{[0]} /\eta|\gg 1$. This consideration results
in the following series in powers of $\eta$,
\begin{eqnarray}
\label{LOadUPPER} \gamma_{0} = -3\ln\biggl( {3^{1/2}}{\eta}\e^{-\gamma_{\rm E}}
\biggr)
- 3 \eta (n+1)
\!\!\!&-&\!\!\! \frac12\eta^2 \biggl( \ft{49}{12} + 8 n + 5 n^2 \biggr)
\nonumber\\
&-&\!\!\!
\ft12 \eta^3
\biggl(
\ft{58}{9} n^3 + \ft{44}{3} n^2 + \ft{335}{18} n + \ft{131}{18}
\biggr)
+
\mathcal{O} (\eta^4)
\, ,
\end{eqnarray}
for one loop, and analogously for two- and three-loop anomalous dimensions
\begin{eqnarray}
\nonumber
\gamma_{1}
\!\!\!&=&\!\!\!
- \ft{1}{12} \pi^2 \gamma_{0}
- \ft94 \zeta (3)
+
\ft14 \eta \left( \sqrt{3}\pi + 6\gamma^{[0]}_0 \right)
+
\ft14 \eta^2
\left(
6 (n + 1) \big( \gamma^{[0]}_0 - 3 \big) + \sqrt{3} \pi (3n + 2)+ \ft{9}{2}
\right)
\\
&+&\!\!\!
\ft14 \eta^3
\left(
- 9 n + \sqrt{3} \pi \left( \ft{59}{12} + 10 n + 7 n^2 \right)
+
2 \left( \ft{49}{12} + 8 n + 5 n^2 \right) \gamma^{[0]}_0
- 33 n^2 - 60 n - \ft{121}{4} \right)
+
\mathcal{O} (\eta^4)
\, ,
\nonumber\\[3mm]
\gamma_{2}
\!\!\!&=&\!\!\!
\ft{11}{720} \pi^4 \gamma_{0}
+ \ft{15}{4} \zeta (5)
+
\ft{1}{8} \pi^2 \zeta (3)
+
\ft18 \eta
\left(
- 27 \zeta (3) - \ft13 \sqrt{3} \pi^2 (\pi + 2\sqrt{3} \gamma^{[0]}_0)\right)
\label{eps_fin}
\\
&+&\!\!\!
\ft18 \eta^2
\Big(
-\ft{5}{4} \pi^2 - 3(\gamma^{[0]}_0)^2-
\sqrt{3} \pi \, \gamma^{[0]}_0 + 18\gamma^{[0]}_0 + 3\sqrt{3}\pi - 27 \zeta(3)
(n+1)
\nonumber\\
&&\qquad\qquad\qquad\qquad\qquad\qquad\quad\ \
+
\ft13 \pi^2 [18 (n+1) - 6\gamma^{[0]}_0 (n+1) - \sqrt{3} \pi (3n+2)]
\Big)
+
\mathcal{O} (\eta^3)
\, , \nonumber
\end{eqnarray}
where $\gamma^{[0]}_0=-3\ln\big( {3^{1/2}}{\eta}\e^{-\gamma_{\rm E}} \big)$. In
Fig.\ \ref{NNLOgammas} we compare the exact spectra at one, two and three loops
with above semiclassical trajectories. In particular, for $N=20$, the
accuracy of the two-loop asymptotic formula for
$\gamma_{1}$
order-by-order in $\eta$ is
\begin{eqnarray}
\label{accur-eps}
\eta: \quad + 0.37\% \, , \qquad\qquad
\eta^2: \quad + 0.11\% \, , \qquad\qquad
\eta^3: \quad + 0.08\% \, .
\end{eqnarray}

We observe that, in each order of perturbation theory, the anomalous dimensions given by
\re{LOadUPPER} and \re{eps_fin} receive corrections $\sim  \ln(1/\eta)$ growing logarithmically with
the conformal spin $\eta^{-1}\sim N$. Retaining the dominant contribution in each order of the
$\eta$-expansion we obtain
\be
\gamma_{N,L=3} = - 3\left[ g^2 \ln \eta + g^4 \lr{-\ft1{12}\pi^2 \ln\eta +\ft32
\eta\ln \eta}+g^6
\lr{\ft{11}{720}\pi^4\ln\eta  - \ft14\eta \ln\eta +\ft{9}{8} \lr{\eta \ln
\eta}^2 }\right]+\ldots
\ee
where ellipses denote corrections suppressed by powers of $\eta\sim 1/N$. It is
straightforward to verify that the same relation can be rewritten as
\begin{align} \notag
\gamma_{N,L=3} &= - 3 \Gamma_{\rm cusp}(g)\left[ \ln\eta + \ft32 \Gamma_{\rm
cusp}(g)\eta\ln \eta+
\ft98 \lr{\Gamma_{\rm cusp}(g)\eta\ln \eta}^2\right]+\ldots
\end{align}
with $\Gamma_{\rm cusp}(g)$ being the cusp anomalous dimension to three loops,
Eq.~\re{GammaCusp}, or equivalently
\be
\label{g-max3}
\gamma_{N,L=3} =  3 \Gamma_{\rm cusp}(g) \ln \lr{N + \ft32 \Gamma_{\rm
cusp}(g)\ln N} + \ldots
\ee
in a perfect agreement with \re{g-max}.

The relation \re{g-max3} suggests that the appearance of $\ln\eta$ terms in the
expansion of the anomalous dimension is an artifact of improper choice of the
expansion parameter. The expansion can be recast in a form free from the large
terms by recalling that in higher orders of perturbation theory the total
conformal spin gets shifted by the anomalous dimension of the Wilson operator
with the corresponding scaling dimension \cite{BelKorMul06}
\begin{eqnarray}
\nonumber
j_0 = N + \ft12 L \qquad\to\qquad j = N + \ft12 L + \ft12 \gamma_{N,L} (g)
\, .
\end{eqnarray}
This implies that in higher orders the anomalous dimension $\gamma_{N,L}$ is
actually a function of the renormalized spin $j$ \cite{BasKor06}, i.e.,
\begin{eqnarray}
\label{symm}
\gamma_{N,L} = f_L (N + \ft12 \gamma_{N,L}) \, .
\end{eqnarray}
This relation can be easily inverted and the first few orders of its
perturbative series read
\begin{eqnarray} \nonumber
f(N)\!\!\!&=&\!\!\!
\gamma(N) - \ft14 (\gamma^2(N))' + \ft{1}{24} (\gamma^3(N))''
+
{\cal O}(\gamma^4(N))
\\
&=&\!\!\!
g^2 \gamma_0(\eta)
+
g^4 \Big[ \gamma_1 (\eta)+
\ft{1}{2}\eta^2 \gamma_0 (\eta) {\gamma_0}'(\eta)
\Big]
\\ \nonumber
&& + g^6 \Big[ \gamma_2 (\eta) + \ft12 \eta^2 (\gamma_0(\eta) \gamma_1(\eta))' +
\ft14 \eta^3
(\gamma_0(\eta))^2 {\gamma_0}' (\eta)
\\ \nonumber
&&\qquad
+ \ft14 \eta^4
\left(
\ft12( \gamma_0(\eta))^2 {\gamma_0}''(\eta)
+
\gamma_0(\eta) ({\gamma_0}'(\eta))^2
\right)
\Big] + {\cal O} (g^8)
\, ,
\label{fN}
\end{eqnarray}
with the expansion coefficient expressed in terms of $\gamma_n$ given in Eqs.\
\re{LOadUPPER} -- \re{eps_fin}.

Since the auxiliary charge $q_2$ depends on the renormalized conformal
Casimir, we may reexpress anomalous dimensions in terms of the parameter
$v = 1/(- q_2(g))^{1/2}$ related to it rather than $\eta$. A straightforward
rearrangement of the semiclassical series yields
\begin{eqnarray}
\gamma_{N,L=3} = g^2 \bar{\gamma}_{0}(v) + g^4 \bar{\gamma}_{1}(v) + g^6
\bar{\gamma}_{2}(v) + {\cal
O}(g^8) \, ,
\end{eqnarray}
where
\begin{eqnarray}\label{LO-v}
\bar{\gamma}_{0}
\!\!\!&=&\!\!\! -3\ln\bigl( {3^{1/2}}{v}\e^{-\gamma_{\rm E}} \bigr) - 3 \left(n+
\ft12\right) v - \ft12
\left( 5 n^2 + 5 n + \ft{10}{3} \right) v^2
\\[1mm]
&&\qquad\qquad\qquad\quad - \ft19 \left( n + \ft12 \right)
\left( 29 n^2 + 29 n + \ft{187}{4} \right) v^3 + {\cal O} (v^4)
\, , \nonumber\\ [1mm]
\label{NLO-v}
\bar{\gamma}_{1}
\!\!\!&=&\!\!\!
-
\ft{1}{12} \pi^2 \bar{\gamma}_{0}
-
\ft{9}{4} \zeta(3)
+
\ft14 \sqrt{3}\pi v
+
\ft34 \left( \sqrt{3} \pi \left( n + \ft12 \right)
+
\ft{3}{2} \right) v^2
\\ [1mm]
&&\qquad\qquad\qquad\quad
+ \ft14
\left( \sqrt{3}\pi \left( 7 n^2 + 7 n + \ft{41}{12} \right)
-
18 \left( n + \ft12 \right) \right) v^3
+ {\cal O}(v^4) \, ,
\nonumber\\ [1mm]
\label{NNLO-v}
\bar{\gamma}_{2}
\!\!\!&=&\!\!\!
\ft{11}{720} \pi^4 \bar{\gamma}_{0}
+
\ft{15}{4} \zeta(5)
+
\ft{1}{8} \pi^2 \zeta(3)
-
\ft{1}{24}\sqrt{3}\pi^3 v
\\ [1mm]
&&\qquad\qquad\qquad\quad
-
\ft18 \left( \ft54 \pi^2 + \sqrt{3}\pi^3 \left( n + \ft12 \right) \right) v^2
+
{\cal O}(v^3)
\, . \nonumber
\end{eqnarray}
As it was mentioned above, $f_L(N)$ is a function of the bare total spin $j_0$ such that the
$N$-dependence of $f_L(N)$ originates solely from the one of the bare charge $q_{2,0} (N)$.
Therefore, re-expanding $f_L(N)$ in the $v_0= 1/(-q_{2,0})^{1/2}$ power series we finally obtain
\be
f_{L=3} (N) = g^2 \bar{f}_0 (v_0) + g^4 \bar{f}_1(v_0) + g^6 \bar{f}_2(v_0) +
{\cal O}(g^8) \, ,
\ee
with $v_0 = [(N+\ft32)(N+\ft12)+\ft34]^{1/2}$ and
\begin{align}\label{LO-f}
\bar{f}_0
&= -3\ln\bigl( {3^{1/2}}{v_0}\e^{-\gamma_{\rm E}} \bigr) - 3 \left(n+
\ft12\right) v - \ft12
\left( 5 \lr{n+\ft12}^2 + \ft{25}{12} \right) v_0^2
\\[1mm] \notag
& \qquad\qquad\qquad\qquad\qquad - \ft19 \left( n + \ft12 \right)
\left( 29 \lr{n+\ft12}^2  + \ft{79}{2} \right) v_0^3 + {\cal O} (v_0^4)
\, ,\\
\label{NLO-f}
\bar{f}_{1}
&=
-
\ft{1}{12} \pi^2 \bar{f}_0
-
\ft{9}{4} \zeta(3)
+
\ft{1}{4} \sqrt{3}\pi v_0
+
\ft34 \left( \sqrt{3} \pi (n + \ft12 )
+
\frac{3}{2} \right) v_0^2
\\
& \qquad\qquad\qquad\qquad\qquad
+ \ft14
\left( \sqrt{3}\pi \left( 7 \lr{n+\ft12}^2  + \ft{5}{3} \right)- 9 (n + \ft12 )
\right) v_0^3
+
{\cal O}(v_0^4) \, , \nonumber\\
\label{NNLO-f}
\bar{f}_{2}
&=
\ft{11}{720} \pi^4 \bar{f}_0
+
\ft{15}{4} \zeta(5)
+
\ft{1}{8} \pi^2 \zeta(3)
-
\ft{1}{24}\sqrt{3}\pi^3 v_0 \\
& \qquad\qquad\qquad\qquad\qquad
-
\ft18 \left( \ft54 \pi^2 + \sqrt{3} \pi^3 (n + \ft12) \right) v_0^2
+
{\cal O}(v_0^3)
\, . \nonumber
\end{align}
We recall that nonnegative integer $n$ enumerates the trajectories close to the upper boundary of
the band \re{twistLband} with the latter corresponding to $n=0$. We observe that, in agreement with
our expectations, the expansion coefficients in front of power suppressed corrections on the
right-hand side of \re{LO-f} -- \re{NNLO-f} are free from logarithmically enhanced contributions.

\section{Discussion and conclusion}

In this paper we analyzed the fine structure of the spectrum of the anomalous
dimensions of Wilson operators in the autonomous $SL(2)$ sector of the maximally
supersymmetric gauge theory in the few lowest orders of the perturbative series
in 't Hooft coupling. We identified two sets of trajectories corresponding to
excited states in the vicinity of the upper and lower boundary of the band
\re{twistLband} which define two distinct analytical continuations for the
anomalous dimensions. Both of them are parameterized by sets of integers
which are related to each other through a set of linear relations. We
focused in great detail on twist-three operators which, while being simple
enough for analytical treatment, contain all salient features involved in
more general considerations of twist-$L$ spectra.

The integrability of the dilatation operator played a key role in our study. We
have applied the method of the Baxter equation, generalized to all orders of
perturbation theory, for the eigenvalues of the Baxter polynomial which
encodes the spectrum of anomalous dimensions of twist operators. Though the
equation is only asymptotic in its nature, i.e., it does not implement
wrapping effects of the putative long-range spin chain underlying the
dilatation operator of the $\mathcal{N} = 4$ superYang-Mills theory, this
difficulty did not show up in the present three-loop considerations. In
the past, the Baxter equation was efficiently used to study asymptotic
regimes in anomalous dimensions at one loop order and was presently
extended to multiloop analyses. The main focus of our consideration was the
construction of a systematic expansion in inverse powers of the Lorentz spin
which define corrections to the logarithmic Sudakov scaling. While the
latter is universal to all gauge theories, the preasymptotic effects are
not and strongly depend on whether the underlying dynamical system is
integrable or not. We developed efficient procedures for systematic studies
of anomalous dimensions in higher orders of perturbation theory generalizing
techniques relying on the Baxter equation used in the past to study
short-range noncompact spin chains in the vicinity of the lower and upper
band of the spectrum. Due to different scaling of the conserved charges
parameterizing the all-order auxiliary transfer matrix entering the
Baxter equation depending on the location within the energy band, we
relied on either asymptotic or semiclassical techniques. Our analysis
yielded a set of analytical formulas which accurately describe the spectra
of anomalous dimension exhibiting very fast convergence of asymptotic and
semiclassical series within the domains of their validity.

The trajectories,  being eigenvalues of certain
hermitian long-range Hamiltonian of the putative AdS/CFT
long-range spin chain with different quantum numbes, do not intersect each other. However, their distribution density
does change as one varies the value of the 't Hooft coupling
constant and other global parameters of Wilson operators. This
raises the question of precise identification of stringy
configurations dual to the excited trajectories at strong coupling
and corresponding dual observables. It is well known that the
lowest trajectory corresponds to the double folded closed string
extended to the boundary of the anti-de Sitter space and spinning
with large angular momentum. The excited trajectories correspond
to the spiky string configuration with its spikes in the vicinity
of the boundary. Along these lines one would gain understanding of
the density of states inside the band for large spins at strong
coupling.

Another problem which emerges from our consideration is the study
of anomalous dimensions of twist-$L$ operators as one moves into
the interior of the band starting from the low boundary. Along the
way, one hits a special set of trajectories \re{GR-K} which scale
at large Lorentz spin as $(K+2) \ln N$. From the point of view of
the spectral curve underlying the magnet this corresponds to
opening double points on the Riemann surface of the auxiliary
problem as one goes upwards the band of eigenvalues. This
continuous deformation of the density of Bethe roots can be
described through the Whitham flow of energy eigenvalues with
respect to the change of the integrals of motion. These issues
will be analyzed in forthcoming publications.

\vspace{0.5cm}

\noindent We would like to thank J. Kota\'nski for collaboration at the early stage and
N. Dorey, A. Rej and M. Staudacher for interesting discussions. This work was
supported by the U.S. National Science Foundation under grant no.\ PHY-0456520 (A.B.), by
the French Agence Nationale de la Recherche under grant ANR-06-BLAN-0142 (G.K. and R.P.)
and by the Bourse du Gouvernement Fran{\c{c}}ais (R.P.). Two of us (A.B. and
R.P.) would like to thank the Institute for Nuclear Theory at University of Washington for its
hospitality and the Department of Energy for partial support during the completion of this work.



\end{document}